\documentclass[entropy,article,authorversion,moreauthors,LaTeX, dvi2pdf,10pt,a4paper,tikz]{mdpi} 
\firstpage{1} 

\Title{Secure and Reliable Key Agreement with Physical Unclonable Functions}

\Author{Onur G\"unl\"u $^{1,4,}$*\orcidA{}, Tasnad Kernetzky $^{2}$\orcidB{}, Onurcan \.{I}\c{s}can $^{3}$\orcidC{}, Vladimir Sidorenko $^{1}$\orcidD{}, Gerhard Kramer $^{1}$\orcidE{}, and Rafael F. Schaefer $^{4}$\orcidF{}}

\AuthorNames{Onur G\"unl\"u, Tasnad Kernetzky, Onurcan \.{I}\c{s}can, Vladimir Sidorenko, Gerhard Kramer, and Rafael F. Schaefer}

\address{%
$^{1}$ \quad Chair of Communications Engineering, Technical University of Munich; \{vladimir.sidorenko, gerhard.kramer\}@tum.de\\
$^{2}$ \quad Associate Professorship of Line Transmission Technology, Technical University of Munich; tasnad.kernetzky@tum.de\\
$^{3}$ \quad Huawei Technologies Duesseldorf GmbH (in 2018); onurcan.iscan@tum.de\\
$^{4}$ \quad Information Theory and Applications Chair, Technische Universit\"at Berlin;\\\hspace*{0.42cm}      \{guenlue, rafael.schaefer\}@tu-berlin.de}
\corres{Correspondence: guenlue@tu-berlin.de}

\firstnote{Parts of this paper were presented at the 2016 IEEE Global Conference on Signal and Information Processing in \cite{ourGlobalSIP} and 2017 IEEE International Conference on Communications in \cite{bizimICC}.} 
 
\abstract{Different transforms used in binding a secret key to correlated physical-identifier outputs are compared. Decorrelation efficiency is the metric used to determine transforms that give highly-uncorrelated outputs. Scalar quantizers are applied to transform outputs to extract uniformly distributed bit sequences to which secret keys are bound. A set of transforms that perform well in terms of the decorrelation efficiency is applied to ring oscillator (RO) outputs to improve the uniqueness and reliability of extracted bit sequences, to reduce the hardware area and information leakage about the key and RO outputs, and to maximize the secret-key length. Low-complexity error-correction codes are proposed to illustrate two complete key-binding systems with perfect secrecy, and better secret-key and privacy-leakage rates than existing methods. A reference hardware implementation is also provided to demonstrate that the transform-coding approach occupies a small hardware area.}

\keyword{key agreement; physical unclonable functions; transform coding; privacy leakage; hardware implementation}

\usetikzlibrary{plotmarks}
\usepackage{pgfplots}
\usetikzlibrary{calc}
\usetikzlibrary{shapes,arrows}
\usetikzlibrary{decorations.markings}
\usetikzlibrary{quotes, math} 
\usetikzlibrary{fit} 
\usetikzlibrary{shapes.multipart} 
\tikzset{
    pics/mux/.default=3/2.5/1.0/0.6/0.1,
    pics/mux/.style    args={#1/#2/#3/#4/#5}
    { code={
        \tikzmath{
            \muxNInps = #1;
            \muxHL = #2;
            \muxW = #3;
            \muxHFrac = #4;
            \muxInpMargin = #5*\muxHL;
            \muxHR = \muxHL*\muxHFrac;
            \muxInpDist = (\muxHL - 2*\muxInpMargin) / (\muxNInps-1);
            \muxNInpsMOne = \muxNInps-1;
        }
        \draw[pic actions]
            (-\muxW/2, \muxHL/2)  coordinate (muxTL) --
            (\muxW/2, \muxHR/2)   coordinate (muxTR) --
            (\muxW/2, -\muxHR/2)  coordinate (muxBR) --
            (-\muxW/2, -\muxHL/2) coordinate (muxBL) --
            cycle;
        \foreach \i in {0,...,\muxNInpsMOne} {
            \coordinate (-in\i) at (-\muxW/2, { \muxHL/2 - \muxInpMargin - \i*\muxInpDist } );
        }
        \coordinate (-out) at (\muxW/2, 0);
        \coordinate (-sel) at ($ (muxBL)!0.5!(muxBR) $);

        \coordinate (-north) at (0,        \muxHL/2);
        \coordinate (-east)  at ( \muxW/2, 0);
        \coordinate (-south) at (0,        -\muxHL/2);
        \coordinate (-west)  at (-\muxW/2, 0);
        \node{\tikzpictext};
    }},
}
\usepackage{subcaption,tabularx}
\newcommand{\Enc}{\mathsf{Enc}}
\newcommand{\Dec}{\mathsf{Dec}}

\newcommand*\xor{\mathbin{\oplus}}
\makeatletter
\newcommand{\wast}{\bBigg@{2}}
\newcommand{\Wast}{\bBigg@{3}}
\newcommand{\vast}{\bBigg@{4}}
\newcommand{\Vast}{\bBigg@{5}}
\makeatother

\makeatletter
\newcommand*\bigcdot{\mathpalette\bigcdot@{.5}}
\newcommand*\bigcdot@[2]{\mathbin{\vcenter{\hbox{\scalebox{#2}{$\m@th#1\bullet$}}}}}
\makeatother

\pgfplotsset{compat=1.13}
\usepackage{dsfont}
\usepackage{array,bm}

\newcommand*\xbar[1]{%
   \hbox{%
     \vbox{%
       \hrule height 0.5pt 
       \kern0.3ex
       \hbox{%
         \kern-0.1em
         \ensuremath{#1}%
         \kern-0.1em
       }%
     }%
   }%
} 

\usepackage{color, colortbl}

\makeatletter
\def\hlinewd#1{%
	\noalign{\ifnum0=`}\fi\hrule \@height #1 %
	\futurelet\reserved@a\@xhline}
\makeatother

\newcolumntype{L}[1]{>{\raggedright\let\newline\\\arraybackslash\hspace{0pt}}m{#1}}
\newcolumntype{C}[1]{>{\centering\let\newline\\\arraybackslash\hspace{0pt}}m{#1}}
\newcolumntype{R}[1]{>{\raggedleft\let\newline\\\arraybackslash\hspace{0pt}}m{#1}}
\begin{document}
\tikzset{XOR/.style={draw,circle,append after command={
				[shorten >=\pgflinewidth, shorten <=\pgflinewidth,]
				(\tikzlastnode.north) edge (\tikzlastnode.south)
				(\tikzlastnode.east) edge (\tikzlastnode.west)
			}
		}
}
\tikzset{line/.style={draw, -latex',shorten <=1bp,shorten >=1bp}}

\section{Introduction}
Secret keys stored in a device can provide intellectual property protection, and device authentication and identification. Non-volatile memory (NVM) is the traditional storage medium for secret keys. Securing the NVM is expensive due to its susceptibility to physical attacks \cite{Suhaegis}. A cheap and safe alternative to the NVM is to use physical identifiers as a source of randomness by applying the concept of \textit{one-way functions} \cite{PappuThesis} to physical systems. 

Invasive (physical) attacks to physical identifiers permanently change the identifier output so that an attacker cannot learn the secret key by using an invasive attack \cite{PappuThesis}. This property eliminates the need for continuous hardware protection \cite{pufintheory}. Physical identifiers like physical unclonable functions (PUFs), e.g., the random start-up value of an uninitialized static random access memory (SRAM) \cite{SRAMPUFFirst} or fine variations of ring oscillator (RO) outputs \cite{ROPUFFirst}, are considered to be random sources with high entropy \cite{GassendThesis}. Thus, we can use PUFs for low-complexity key storage in, e.g., internet of things (IoT) applications like securing a surgical robot against hacking.

There are multiple \textit{key-generation}, or generated-secret (GS), and \textit{key-binding}, or chosen-secret (CS), methods to reconstruct secret keys from noisy PUF outputs, where the key is generated from the PUF outputs or bound to them, respectively. Code-offset fuzzy extractors \cite{FuzzyBasic} are examples of key-generation methods and the \text{fuzzy commitment} scheme \cite{juelsfuzzy} is a key-binding method. Code constructions based on Wyner-Ziv (WZ) coding are illustrated in \cite{bizimWZ} to asymptotically achieve the information-theoretic limits for the GS and CS models. These constructions might have high complexity, which is undesired for, e.g., IoT applications. In addition, since a key should be stored in a secure database for both models, it is more practical to allow a trusted entity to choose the secret key bound to a PUF output. Thus, in this paper, we aim at further improving reliability, privacy, secrecy, and hardware cost performance of a transform-coding algorithm, explained next, that is applied to PUF outputs in combination with the fuzzy commitment scheme.

PUFs have similar features to biometric identifiers like fingerprints. Both identifier types have correlated and noisy outputs due to surrounding environmental conditions \cite{RoelMaesbook}. Correlation in PUF outputs leaks information about the secret key, which causes \textit{secrecy leakage}, and about the PUF output, causing \textit{privacy leakage} \cite{IgnatenkoTrans,Lai10,bizimproofarxiv}. Moreover, noise reduces reliability of PUF outputs and error-correction codes are needed to satisfy the reliability requirements. The transform-coding approach \cite{benimthesis,bizimpaper} in combination with a set of scalar quantizers has made its way into secret-key binding with continuous-output biometric and physical identifiers, as they allow reducing the output correlation and adjusting the effective noise at the PUF output. For instance, the discrete cosine transform (DCT) is the building block in \cite{bizimpaper} to generate a uniformly distributed bit sequence from RO outputs under varying environmental conditions. Efficient post-processing steps are applied to obtain more reliable PUF outputs rather than changing the hardware architecture, so standard components can be used. This transform-coding approach improves on the existing approaches in terms of the reliability under varying environmental conditions and maximum key length \cite{bizimpaper, bizimtemperature}. We apply this algorithm to PUF outputs with further significant improvements by designing the transformation and error-correction steps jointly.

Information-theoretic limits for the fuzzy commitment scheme are given in \cite{IgnaFuzzy}. We use these information-theoretic limits to compare error-correction codes proposed for the transform-coding algorithm with the limits. Similar analyses were conducted for biometric identifiers in \cite{ignatenko2014privacy}, but their assumptions such as independent and identically distributed (i.i.d.) identifier outputs and maximum block-error probability constraint $P_B\!=\! 10^{-2}$ are not realistic. We therefore consider highly correlated RO outputs with the constraint $P_B\!\leq\!\displaystyle 10^{-9}$, which are realistic for security applications that use PUFs \cite{Pufky}. 

\subsection{Summary of Contributions and Organization}
We improve the DCT-based algorithm of \cite{bizimpaper} by using different transforms and reliability metrics. We also propose error-correction codes that achieve better (secret-key, privacy-leakage) rate tuples than previous code designs. A summary of the main contributions is as follows.
\begin{itemize}
	\item We compare a set of transforms to improve the performance of the transform coding algorithm in terms of the maximum secret-key length, decorrelation efficiency, uniqueness and security of the extracted bit sequence, and computational complexity. 
	\item Two quantization methods with different reliability metrics are proposed to address multiple design objectives for PUFs. One method aims at maximizing the length of the bit sequence extracted from a fixed number of ROs, whereas the second method provides reliability guarantees for each output in the transform domain by fixing the decoding capability of a decoder used for error correction. 
	\item We give a reference hardware design for the transform with the smallest computational complexity, among the set of transforms considered, in combination with the second quantization method to illustrate that our algorithm occupies a small hardware area. Our results are comparable to hardware area results of previous RO PUF designs.
	\item Error-correction codes that satisfy the block-error probability constraints for practical PUF systems are proposed for both quantization methods to illustrate complete key-binding systems with perfect secrecy. The proposed codes operate at better rate tuples than previously proposed codes for the fuzzy commitment scheme. Our quantizer designs also allow us to significantly reduce the gap to the optimal (secret-key, privacy-leakage) rate point achieved by the fuzzy commitment scheme.	   
\end{itemize}

This paper is organized as follows. In Section~\ref{sec:systemmodel}, we define the fuzzy commitment scheme that uses PUF outputs as the randomness source. The transform-coding algorithm proposed to extract a reliable bit sequence from RO PUFs is explained in Section~\ref{sec:commonsteps}. We propose two different quantization methods with different reliability metrics in Section~\ref{sec:quantandcodedesign}. In Section~\ref{sec:comparisons}, we illustrate the small hardware area of the proposed algorithm with a reference hardware design, and the gains in terms of reliability, security, and maximum secret-key length as compared to the existing methods. Our proposed error-correction codes, and their secrecy and privacy performance are described in Section~\ref{sec:correction}. Section~\ref{sec:conclusion} concludes the paper.

\subsection{Notation}
Upper case letters represent random variables and lower case letters their realizations. A letter with superscript denotes a string of variables, e.g., $\displaystyle X^N\!=\!X_1\ldots X_i\ldots X_N$, and a subscript denotes the position of a variable in the string. A random variable $\displaystyle X$ has probability mass $\displaystyle P_X$ or probability density $f_X$. Calligraphic letters such as $\displaystyle \mathcal{X}$ denote sets, and set sizes are denoted as $\displaystyle |\mathcal{X}|$. Bold letters such as $\mathbf{H}$ represent matrices. $\Enc(\cdot)$ is an encoder mapping and $\Dec(\cdot)$ is a decoder mapping. $X-Y-Z$ indicates a Markov chain. $H_b(x)=-x\log_2 x- (1-x)\log_2 (1-x)$ is the binary entropy function. The $*$-operator is defined as $\displaystyle p*x = p(1-x)+(1-p)x$. The operator $\xor$ represents the element-wise modulo-2 summation. A binary symmetric channel (BSC) with crossover probability $p$ is denoted by BSC($p$). $X^n\sim\text{Bern}^n(\alpha)$ denotes that $X^n$ is an i.i.d. binary sequence of random variables with $\Pr[X_i=1]=\alpha$ for $i=1,2,\ldots,n$. $\text{Unif}\,[1\!:\!|\mathcal{X}|]$ represents a uniform distribution over the integers from $1$ to $|\mathcal{X}|$. A linear error-correction code $\mathcal{C}$ with parameters $(n,k,d)$ has block length $n$, dimension $k$, and minimum distance $d$ so that it can correct up to $\lfloor \frac{d-1}{2}\rfloor$ errors.
.

\section{System Model and the Fuzzy Commitment Scheme}\label{sec:systemmodel}
Consider a RO as a source that generates a symbol $\tilde{x}$. Systematic variations in RO outputs in a two-dimensional array are less than the systematic variations in one-dimensional ROs \cite{maiti2011improved}. We thus consider a two-dimensional RO array of size $\displaystyle L\!= r\!\times\! c$ and represent the array as a vector random variable $\widetilde{X}^L$. Suppose there is a single \emph{PUF circuit}, i.e., a single two-dimensional RO array, in each device with the same circuit design, and it emits an output $\displaystyle \widetilde{X}^L$ according to a probability density $f_{\widetilde{X}^L}$. Each RO output is disturbed by mutually-independent additive white Gaussian noise (AWGN) and the vector noise is denoted as $\widetilde{Z}^L$. Define the noisy RO outputs as $\widetilde{Y}^L\! =\! \widetilde{X}^L \!+\! \widetilde{Z}^L$. Observe that $\widetilde{X}^L$ and $\widetilde{Y}^L$ are correlated. A secret key can thus be agreed by using these outputs of the same RO array \cite{AhlswedeCsiz,Maurer,IgnatenkoTrans,Lai10}. 

One needs to extract random sequences with i.i.d. symbols from $\widetilde{X}^L$ and $\widetilde{Y}^L$ to employ available information-theoretic results for secret-key binding with identifiers. We propose an algorithm that extracts nearly i.i.d. binary and uniformly distributed random vectors $\displaystyle \left(X^N\!,\,Y^N\right)$ from $\widetilde{X}^L$ and $\widetilde{Y}^L$, respectively. For such $\displaystyle X^N$ and $\displaystyle Y^N$, we can define a binary error vector as $E^N\! =\! X^N\! \xor\! Y^N$. The random sequence $\displaystyle E^N$ corresponds to a sequence of i.i.d. Bernoulli random variables with parameter $p$, i.e., $E^N\sim\text{Bern}^n(p)$. The channel $P_{Y|X}$ is thus a BSC$(p)$. 

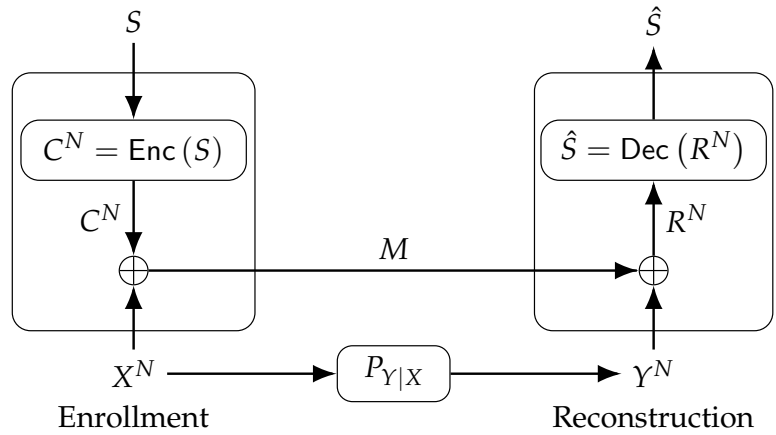
\begin{figure}
	\centering
	\resizebox{0.655\linewidth}{!}{
		
		\begin{tikzpicture}
		\node (a) at (0,-1.5) [XOR,scale=1.0] {};
		\node (b) at (6,-1.5) [XOR,scale=1.0] {};
		\node (f) at (0,-0.7) [draw,rounded corners = 6pt, minimum width=2.8cm,minimum height=3cm, align=left] {};
		\node (g) at (6,-0.7) [draw,rounded corners = 6pt, minimum width=2.75cm,minimum height=3cm, align=left] {};
		\node (d) at (0,-0.1) [draw,rounded corners = 6pt, minimum width=2.6cm,minimum height=0.7cm, align=left] {$
			C^N = \Enc\left(S\right)$};
		\node (c) at (3,-2.7) [draw,rounded corners = 5pt, minimum width=1.3cm,minimum height=0.65cm, align=left] {$P_{Y|X}$};
		\node (e) at (6,-0.1) [draw,rounded corners = 6pt, minimum width=2.6cm,minimum height=0.7cm, align=left] {$\hat{S} = \Dec\left(R^N\right)$};
		\draw[decoration={markings,mark=at position 1 with {\arrow[scale=1.5]{latex}}},
		postaction={decorate}, thick, shorten >=0.6pt] (a.east) -- (b.west) node [midway, above] {$M$};
		\node (a1) [below of = a, node distance = 1.2cm] {$X^N$};
		\node (b1) [below of = b, node distance = 1.2cm] {$Y^N$};
		\draw[decoration={markings,mark=at position 1 with {\arrow[scale=1.5]{latex}}},
		postaction={decorate}, thick, shorten >=1.4pt] (a1.north) -- (a.south);
		\draw[decoration={markings,mark=at position 1 with {\arrow[scale=1.5]{latex}}},
		postaction={decorate}, thick, shorten >=1.4pt] (a1.east) -- (c.west);
		\draw[decoration={markings,mark=at position 1 with {\arrow[scale=1.5]{latex}}},
		postaction={decorate}, thick, shorten >=1.4pt] (c.east) -- (b1.west);
		\draw[decoration={markings,mark=at position 1 with {\arrow[scale=1.5]{latex}}},
		postaction={decorate}, thick, shorten >=1.4pt] (b1.north) -- (b.south);
		\node (a2) [above of = d, node distance = 1.5cm] {$S$};
		\node (f2) [below of = f, node distance = 2.5cm] {Enrollment};
		\node (g2) [below of = g, node distance = 2.5cm] {Reconstruction};
		\node (b2) [above of = e, node distance = 1.5cm] {$\hat{S}$};
		\draw[decoration={markings,mark=at position 1 with {\arrow[scale=1.5]{latex}}},
		postaction={decorate}, thick, shorten >=1.4pt] (e.north) -- (b2.south);
		\draw[decoration={markings,mark=at position 1 with {\arrow[scale=1.5]{latex}}},
		postaction={decorate}, thick, shorten >=1.4pt] (a2.south) -- (d.north); 
		\draw[decoration={markings,mark=at position 1 with {\arrow[scale=1.5]{latex}}},
		postaction={decorate}, thick, shorten >=1.4pt] (b.north) -- (e.south) node [midway, right] {$R^N$};
		\draw[decoration={markings,mark=at position 1 with {\arrow[scale=1.5]{latex}}},
		postaction={decorate}, thick, shorten >=1.4pt]  (d.south) -- (a.north) node [midway, left] {$C^N$};;
		\end{tikzpicture}
	}
	\caption{The fuzzy commitment scheme.}\label{fig:fuzzycommitment}
\end{figure}

The fuzzy commitment scheme reconstructs a secret key by using correlated random variables without leaking any information about the secret key \cite{juelsfuzzy}. The fuzzy commitment scheme is depicted in Fig.~\ref{fig:fuzzycommitment}, where an encoder $\Enc$  embeds a secret key, uniformly distributed according to $\text{Unif}\,[1\!:\!|\mathcal{S}|]$, into a binary codeword $\displaystyle C^N$ that is added modulo-2 to the binary PUF-output sequence $\displaystyle X^N$ during enrollment. The resulting sequence is the public helper data $\displaystyle M$, which is sent through an authenticated and noiseless channel. The modulo-2 sum of the helper data $M$ and $Y^N$ gives the result 
\begin{align}
R^N&=M\xor Y^N=C^N\!\xor E^N
\end{align}
which is later mapped to an estimate $\displaystyle \hat{S}$ of the secret key by the decoder $\displaystyle \Dec$ during reconstruction.

\begin{Definition}\label{def:ratepair}
A secret-key vs. privacy-leakage rate pair $\displaystyle \left(R_s\!\;\text{,}\;\!R_l\right)$ is achievable by the fuzzy commitment scheme with perfect secrecy, i.e., zero secrecy leakage, if, given any $\epsilon\!>\!0$, there is some $N\!\geq\!1$ and an encoder and decoder for which $\displaystyle R_s=\frac{\log_2|\mathcal{S}|}{N}$ and
\begin{alignat}{2}
&\Pr[S\ne\hat{S}] \leq \epsilon && (\text{reliability}) \label{eq:reliabilityconst}\\
&I\left(S;M\right)\!=\!0 && (\text{perfect secrecy})\label{eq:secrecyconst}\\
&\frac{1}{N}I\left(X^N;M\right) \leq R_l+\epsilon \quad\quad\quad&&(\text{privacy})  \label{eq:privacyconst}.
\end{alignat}
\end{Definition}

\begin{Theorem}[\cite{IgnaFuzzy}]
The achievable secret-key vs. privacy-leakage rate region for the fuzzy commitment scheme with a channel $P_{Y|X}$ that is a BSC$(p)$, uniformly distributed $X$ and $Y$, and zero secrecy leakage is 
\begin{align}
\mathcal{R}\! =\! \{&\left(R_s,R_l\right)\!\colon\! 0\leq R_s\leq 1-H_b(p),\quad R_l\geq 1\!-\!R_s\}\label{eq:ls0}.
\end{align}
\end{Theorem}

The region $\mathcal{R}$ suggests that any (secret-key, privacy-leakage) rate pair that sums up to $1$ bit/source-bit is achievable with the constraint that the secret-key rate is at most the channel capacity of the BSC. Furthermore, smaller secret-key rates and greater privacy-leakage rates than these rates are also achievable. 

The fuzzy commitment scheme is a particular realization of the CS model. The region $\mathcal{R}_{\text{cs}}$ of all achievable (secret-key, privacy-leakage) rate pairs for the CS model with a negligible secrecy-leakage rate, where a generic encoder is used to confidentially transmit an embedded secret key to a decoder that observes $Y^N$ and the helper data $M$, is given in \cite{IgnatenkoTrans} as
\begin{align}
\mathcal{R}_{\text{cs}}\! =\! &\bigcup_{P_{U|X}}\!\Bigg\{\left(R_s,R_l\right)\!\colon\!\quad 0\leq R_s\leq I(U;Y),\;\;\;R_l\geq I(U;X)-I(U;Y)\Bigg\}\label{eq:chosensecret}
\end{align}
where $U-X-Y$ forms a Markov chain and the alphabet $\mathcal{U}$ of the auxiliary random variable $U$ can be limited to have the size $\displaystyle |\mathcal{U}|\!\leq\!|\mathcal{X}|+1$. The fuzzy commitment scheme is optimal, i.e., it achieves a boundary point of $\mathcal{R}_{\text{cs}}$, for a BSC $P_{Y|X}$ with crossover probability $p$, only at the point $\displaystyle (R_s^*,R_l^*)\!=\!(1\!-\!H_b(p),H_b(p))$ \cite{IgnaFuzzy}. This point corresponds to the highest achievable secret-key rate. Note that the region $\mathcal{R}_{\text{cs}}$ gives an outer bound for the perfect-secrecy case (see \cite{IgnatenkoTrans} for discussions).

\section{Transform Coding Steps}\label{sec:commonsteps}
\begin{figure}[t]
	\centering
	\includegraphics[width=0.735\textwidth, height=0.785\textheight, keepaspectratio]{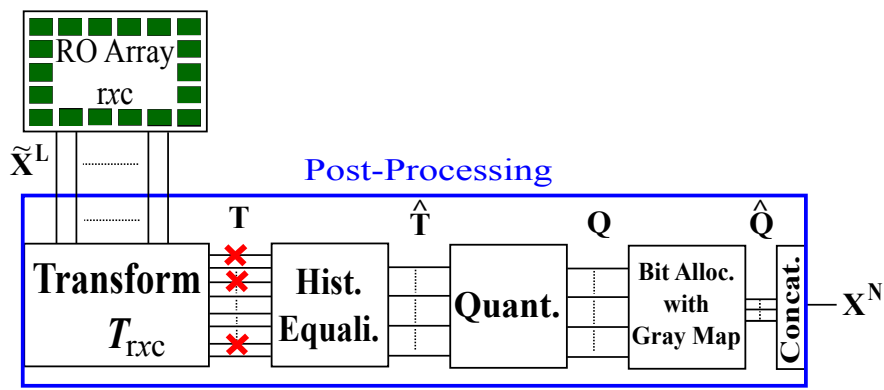}
	\caption{Transform-coding steps.} 
	\label{fig:postprocessing}
\end{figure}

The aim of transform coding is to reduce the correlations between RO outputs by using a linear transformation. We propose a transform-coding algorithm that extends the work in \cite{benimthesis} and \cite{bizimpaper}. Optimizations of the quantization and error-correction parameters to maximize the security and reliability performance, and a simple method to decrease storage are its main steps. The output of these post-processing steps is a bit sequence $X^N$ (or its noisy version $Y^N$) used in the fuzzy commitment scheme. We consider the same post-processing steps for the enrollment and reconstruction with the exception that during enrollment the design parameters are determined by the device manufacturer depending on the source statistics. It thus suffices to discuss only the enrollment steps. Fig.~\ref{fig:postprocessing} shows the post-processing steps that include transformation, histogram equalization, quantization, bit assignment, and bit-sequence concatenation.

RO outputs $\widetilde{X}^L$ in an array are correlated due to, e.g., the surrounding logic \cite{ROAnalysis}. A transform $\emph{T}_{r\!\times\!c}(\cdot)$ of size $\displaystyle r\!\times\! c$ is applied to an array of RO outputs to reduce correlations. Decorrelation performance of a transform depends on the source statistics. We model each output $T$ in the transform domain, called \emph{transform coefficient}, obtained from a RO-output dataset in \cite{ROLarge} by using the corrected Akaike information criterion (AICc) \cite{CorrAIC} and the Bayesian information criterion (BIC) \cite{BIC2}. These criteria suggest that a Gaussian distribution can be fitted to each transform coefficient $T$ for the discrete cosine transform (DCT), discrete Walsh-Hadamard transform (DWHT), discrete Haar transform (DHT), and Karhunen-Lo\`eve transform (KLT), which are common transforms considered in the literature for image processing, digital watermarking, etc. \cite{mySPbook}. We use maximum-likelihood estimation \cite{MLE} to derive unbiased estimates for the parameters of Gaussian distributions. 

The histogram equalization step in Fig.~\ref{fig:postprocessing} converts the probability density of the $i$-th coefficient $T_i$ into a standard normal distribution such that $ \widehat{T}_i = \frac{T_i - \mu_i}{\sigma_i}$, where $\displaystyle \mu_i$ is the mean and $\displaystyle \sigma_i$ is the standard deviation of the $i$-th transform coefficient for all $\displaystyle i\!=\!1,2,\ldots,L$. Quantization steps for all transform coefficients are thus the same. Without histogram equalization, we need a different quantizer for each transform coefficient. Therefore, the histogram equalization step reduces the storage for the quantization steps. Transformed and equalized coefficients $\displaystyle \widehat{T}_i$ are independent if the transform $\emph{T}_{r\!\times\!c}(\cdot)$ decorrelates the RO outputs perfectly and the transform coefficients $\displaystyle T_i$ are jointly Gaussian. One can thus use a scalar quantizer for all coefficients without a performance loss. We propose scalar quantizer and bit extraction methods that satisfy the security and reliability requirements of the fuzzy commitment scheme with the independence assumption, in combination with a correlation-thresholding approach, as discussed below.

\section{Quantizer and Code Designs}\label{sec:quantandcodedesign}
The aim of the post-processing steps in Fig.~\ref{fig:postprocessing} is to extract a uniformly-random bit sequence $\displaystyle X^N$. We use a quantizer $\displaystyle Q(\cdot)$ with quantization-interval values $\displaystyle k=1,2,\cdots,2^{K_i}$, where $\displaystyle K_i$ is the number of bits we extract from the $i$-th coefficient $\widehat{T}_i$ for $i\!=\!1,2,\ldots,L$. We have
\begin{align}
Q(\hat{t}_i) = k\quad \text{if}\quad  b_{k-1}\!<\!\hat{t}_i\!\leq\!b_k \label{eq:quantizer}
\end{align}
and we choose $\displaystyle b_k = \Phi^{-1}\left(\frac{k}{2^{K_i}}\right)$, where $\displaystyle \Phi^{-1}(\cdot)$ is the quantile function of the standard normal distribution. The quantizer output $k$ is assigned to a bit sequence of length $K_i$. The chosen permutation of assigned bit sequences does not affect the security performance. However, the most likely error event when we quantize $\widehat{T}_i$ is a jump to a neighboring quantization step due to zero-mean noise. We thus apply a Gray mapping when we assign bit sequences of length $K_i$ to the integers $k=1,2,\ldots,2^{K_i}$ so that neighboring bit sequences change only in one bit position.

We next propose two different reliability metrics for joint quantizer and code designs. The first metric results in BSC measurements of each extracted bit with approximately the same crossover probability. This method extracts a different number of bits from each transform coefficient. The code design is then done for a fixed crossover probability of the BSCs. The second method fixes the maximum number of erroneous transform coefficients and considers an error-correction code that can correct all error patterns with up to a fixed number of errors. 
\subsection{Quantizer Design with Fixed Measurement Channels}\label{subsec:quant1}
Observe that with the quantizer in (\ref{eq:quantizer}) and a Gray mapping, one can model the channel between a bit extracted from the enrollment outputs $\displaystyle \widetilde{X}^L$ and the corresponding bit extracted from the reconstruction outputs $\displaystyle \widetilde{Y}^L$ as a BSC with a fixed average crossover probability $p_b$. Our algorithm thus fixes an average crossover probability $p_b$ such that the error-correction step in the fuzzy commitment scheme can satisfy the maximum block-error probability of $\displaystyle 10^{-9}$. The algorithm enforces that each output $\displaystyle \hat{t}_i$ results in an average bit error probability as close as possible to, but not greater than, $p_b$ by adapting the number of bits $\displaystyle K_i(p_b)$ extracted from the $i$-th coefficient $\widehat{T}_i$ for all $\displaystyle i\!=\!1,2,\ldots,L$. We use the \emph{average fractional Hamming distance} $D(K)$ between the quantization intervals assigned to the original and noisy coefficients as a metric to determine $\displaystyle K_i(p_b)$. Define
\begin{align}
D_i(K)\!=\!\frac{1}{K}\int_{-\infty}^\infty\! \int_{-\infty}^\infty \!\Bigg(\! \sum\limits_{k=1}^{2^K}\!\Pr[Q(\hat{t}\!+\!\hat{n})=k] {\mathrm{HD}}_{k}(\hat{t})\!\Bigg)\cdot f_{\widehat{T}_i}(\hat{t})f_{\widehat{N}_i}(\hat{n})\mathrm{d}\hat{t}\mathrm{d}\hat{n}\label{eq:meancost}
\end{align}
where ${\mathrm{HD}}_{k}(\hat{t})$ is the Hamming distance between the bit sequences assigned to the $k$-th quantization interval and to the interval $Q(\hat{t})$, and $\widehat{N}_i$ represents the Gaussian noise in the $i$-th coefficient after histogram equalization. We then determine $K_i(p_b)$ as the greatest number of bits $K$ such that $D_i(K)\!\leq\!p_b$.

The first coefficient, i.e., DC coefficient, $\widehat{T}_1$ is not used since its value is a scaled version of the mean of the RO outputs in the array, which is generally known by an eavesdropper. Ambient-temperature and supply-voltage variations have a highly-linear effect on the RO outputs, so the DC coefficient is the most affected coefficient, which is another reason not to use the DC coefficient \cite{bizimtemperature}. Therefore, the total number $N(p_b)$ of extracted bits from all transform coefficients for a fixed $p_b$ is 
\begin{align}
N(p_b) = \sum_{i=2}^{L} K_i(p_b)\label{eq:Lpb}.
\end{align}
We calculate the maximum secret-key length $S_{\text{max}}$ by using (\ref{eq:ls0}) for a BSC$(p_b)$ with the maximum secret-key rate $R_s^*\!=\!1\!-\!H_b(p_b)$ as
\begin{align}
S_{\text{max}} = (1-H_b(p_b))\cdot N(p_b)\label{eq:smax}
\end{align}
which is used to compare different transforms and to decide whether one can use an RO PUF with fixed number of ROs and $p_b$ for secret-key binding. For instance, for the advanced encryption standard (AES), the minimum secret-key length is 128 bits. However, the rate region $\mathcal{R}$ in (\ref{eq:ls0}) is valid for large $N$. One thus needs to consider the rate loss due to a finite block length for a system design. 

\subsection{Quantizer Design with Fixed Number of Errors}\label{subsec:quant2}
We now propose a \emph{conservative} approach, based on the assumption that either all bits extracted from a transform coefficient are correct or they all flip, to provide reliability guarantees. The correctness probability $P_c$ of a transform coefficient is defined to be the probability that all bits associated with this coefficient are correct. We use this metric to determine the number of bits extracted from each coefficient such that there is an encoder and a bounded minimum distance decoder (BMDD) that satisfy the block-error probability constraint $P_B\!\leq\!10^{-9}$. This approach results in reliability guarantees for the random-output RO arrays. 

For a $K$-bit quantizer and the quantization boundaries $b_k$ as in (\ref{eq:quantizer}) for an equalized (i.e., standard) Gaussian transform coefficient $\hat{T}$, we obtain the correctness probability 
\begin{align}
P_c&(K) \!=\!\sum_{k=0}^{2^K-1}\!\int\displaylimits_{b_{k}}^{b_{k+1}}\Bigg[\!Q\Big(\frac{b_{k}\!-\!\hat{t}}{\sigma_{\hat{n}}}\Big)\!-\!Q\Big(\frac{b_{k+1}\!-\!\hat{t}}{\sigma_{\hat{n}}}\Big)\!\Bigg]f_{\widehat{T}}(\hat{t})d{\hat{t}}\label{eq:correctness}
\end{align}
where $\displaystyle \sigma^2_{\hat{n}}$ is the noise variance and $\displaystyle f_{\widehat{T}}$ is the probability density of the standard Gaussian distribution. 
 
Suppose our channel decoder can correct all errors in up to $\displaystyle C_{\text{max}}$ transform coefficients. Suppose further that coefficient errors occur independently and that the correctness probability $\displaystyle P_{c,i}(K)$ of the $i$-th coefficient $\widehat{T}_i$ for $i\!=\!1,2,\ldots,L$ is at least $\displaystyle \xbar{P}_c(C_{\text{max}})$. A sufficient condition for satisfying the block-error probability constraint $P_B\!\leq\!10^{-9}$ is that $\displaystyle \xbar{P}_c(C_{\text{max}})$ satisfies the inequality
\begin{align}
 \sum_{c=C_{\text{max}}+1}^{L}{L \choose c}{(1\!-\!\xbar{P}_c(C_{\text{max}}))}^{c}{\xbar{P}_c(C_{\text{max}})}^{L-c}\!\leq\! 10^{-9}\label{eq:threshold}.
\end{align} 
We thus determine the number $K_i$ of bits extracted from the $i$-th transform coefficient as the maximum value $K$ such that $\displaystyle P_{c,i}(K)\geq \bar{P}_{c}(C_{\text{max}})$. Similar to Section~\ref{subsec:quant1}, we choose $K_1\!=\!0$ so that the total number $\displaystyle N(C_{\text{max}})$ of extracted bits is
\begin{align}
 N(C_{\text{max}})\!=\!\sum_{i=2}^L K_i \label{eq:totalbits}.
\end{align}
In the worst case, the coefficients in error are the coefficients from which the greatest number of bits is extracted. We sort the numbers $K_i$ of bits extracted from all coefficients in descending order such that $\displaystyle K^{\prime}_{i}\!\geq\!K^{\prime}_{i+1}$ for all $i\!=\!1,2,\ldots,L-1$. The channel decoder thus must be able to correct up to
\begin{align}
 e( C_{\text{max}}) = \sum_{i=1}^{C_{\text{max}}}  K^{\prime}_{i}\label{eq:minimume}
\end{align}
bit errors, which can be satisfied by using a block code with minimum distance $\displaystyle d_{\text{min}}\!\geq\!2e(C_{\text{max}})\!+\!1$. 

Suppose a key bound to physical identifiers in a device is used in the AES with a uniformly-distributed secret-key with a length of 128 bits. The block code used in the fuzzy commitment scheme should thus have a code length of at most $\displaystyle N(C_{\text{max}})$ bits, code dimension of at least $128$ bits, and minimum distance of $\displaystyle d_{\text{min}}\!\geq\!2e(C_{\text{max}})+1$ for a fixed $\displaystyle C_{\text{max}}$. The code rate should be as high as possible to operate close to the optimal (secret-key, privacy-leakage) rate point of the fuzzy commitment scheme. This optimization problem is hard to solve. We illustrate by an exhaustive search over a set of $\displaystyle C_{\text{max}}$ values and over a selection of algebraic codes that there is a channel code that satisfies these constraints with a reliability guarantee for each extracted bit. Restricting our search to codes that admit low-complexity encoders and decoders is desired for IoT applications, for which complexity is the bottleneck.

Note that the listed conditions are conservative. For a given transform coefficient, the correctness probability can be significantly greater than the correctness threshold $\displaystyle \xbar{P}_c(C_{\text{max}})$. Secondly, due to Gray mapping, it is more likely that less than $K_i$ bits are in error when the $i$-th coefficient is erroneous. Thirdly, it is also unlikely that the bit errors always occur in the transform coefficients from which the greatest number of bits is extracted. Therefore, even if a channel code cannot correct all error patterns with up to $e( C_{\text{max}})$ errors, it can still be the case that the block-error probability constraint is satisfied. We illustrate such a case in the next section.
\section{Performance Evaluations}\label{sec:comparisons}
Suppose the device output $\widetilde{X}^L$ is a vector random variable with the autocovariance matrix $\mathbf{C_{\widetilde{X}\widetilde{X}}}$. Consider RO arrays of sizes $8\!\times \!8$ and $16\!\times \!16$. Autocovariance matrix elements of such RO array outputs and noise are estimated from the dataset in \cite{ROLarge}. We compare the DCT, DWHT, DHT, and KLT in terms of their decorrelation efficiency, maximum secret-key length, complexity, uniqueness, and security. 


\subsection{Decorrelation Performance}
One should eliminate correlations between the RO outputs and make them independent to extract uniform bit sequences by treating each transform coefficient separately. We use the \textit{decorrelation} \textit{efficiency} $\displaystyle \eta_c$ \cite{decorrelation} as a decorrelation performance metric. Consider the autocovariance matrix $\mathbf{C_{TT}}$ of the transform coefficients, so $\displaystyle \eta_c$ of a transform is 
\begin{align}
\eta_c = 1-\frac{\sum\limits_{a=0}^{L}\sum\limits_{b=0}^{L}|\mathbf{C_{TT}}(a,b)|\mathds{1}\{a\!\ne\!b\}}{\sum\limits_{a=0}^{L}\sum\limits_{b=0}^{L}|\mathbf{C_{\widetilde{X}\widetilde{X}}}(a,b)|\mathds{1}\{a\!\ne\!b\}}
\end{align}
where the indicator function $\displaystyle \mathds{1}\{a\!\ne\!b\}$ takes on the value 1 if $\displaystyle a\!\ne\!b$ and 0 otherwise. The decorrelation efficiency of the KLT is 1, which is optimal \cite{decorrelation}. We list the average decorrelation efficiency results of other transforms in Table~\ref{tab:decorrelationeff}. All transforms have similar and good decorrelation efficiency performance for the RO outputs in the dataset in \cite{ROLarge}. The DCT and DHT have the highest efficiency for $\displaystyle 8\!\times \!8$ RO arrays, whereas for $\displaystyle 16\!\times \!16$ RO arrays, the best transform is the DWHT. Table~\ref{tab:decorrelationeff} indicates that increasing the array size improves $\displaystyle \eta_c$.

\begin{table}[t]
	\caption{The average RO output decorrelation-efficiency results.}
	\centering
	\begin{tabular}{ c|C{1.7cm}|C{1.7cm}|C{1.7cm}| }
		\cline{2-4}
		& DCT & DWHT & DHT\\
		\cline{1-4}
		\multicolumn{1}{|c|}{$\displaystyle \eta_c$ for $8\times 8$} &  0.9978 &  0.9977&  0.9978\\
		\cline{1-4}
		\multicolumn{1}{|c|}{$\displaystyle \eta_c$ for $16\times 16$} & 0.9987 &0.9988 &  0.9986\\
		\cline{1-4}
	\end{tabular}\label{tab:decorrelationeff}
\end{table}

\subsection{Maximum Secret-key Length}
The maximum number of bits extracted with the method given in Section~\ref{subsec:quant2} depends on the fixed number of transform coefficients that are in error. Moreover, the method uses a conservative metric. However, for the method given in Section~\ref{subsec:quant1}, we can optimize the number of bits extracted from each coefficient to maximize the secret-key length. We therefore consider only the method in Section~\ref{subsec:quant1} for maximum key-length comparisons.

The secret key $S$ should satisfy the length constraints of the cryptographic primitives that use it. Consider again the AES with a 128-bit secret key. We compare different transforms by calculating the maximum secret-key lengths $S_{\text{max}}$, defined in (\ref{eq:smax}), for various crossover probabilities $p_b$ that can be obtained by applying the post-processing steps in Fig.~\ref{fig:postprocessing}. For RO array dimensions $\displaystyle 8\!\times \!8$, we show $\displaystyle S_{\text{max}}$ results of the considered transforms in Fig.~\ref{fig:maxsecretkey}. For $p_b\!\leq\!0.05$, $R_s^*$ is high but $N(p_b)$ is small, so $\displaystyle S_{\text{max}}$ is mainly determined by $N(p_b)$, as depicted in Fig.~\ref{fig:maxsecretkey}. For $p_b\!\geq\!0.07$, $N(p_b)$ is high but $R_s^*$ mainly determines $\displaystyle S_{\text{max}}$, which is small.

The DHT, DWHT, and DCT have similar $S_{\text{max}}$ results and the KLT has worse performance than the others, which is mainly determined by the signal-to-noise ratio (SNR) in the transform domain. This illustrates that a transform's $\displaystyle \eta_c$ performance for the estimated RO output distribution and its $S_{\text{max}}$ performance for the estimated RO output and noise distributions can be different. We determine a crossover probability range $\displaystyle \mathcal{P}\!=\![0.05,0.07]$ such that the secret-key length of all transforms are close to their maximum and greater than 128. For a BSC with crossover probability $\displaystyle p\!\in\!\mathcal{P}$, we design error-correction codes such that $P_B\!\leq\!\displaystyle 10^{-9}$ is satisfied. The crossover probability range considered in \cite{Pufky} is $[0.12, 0.14]$, while $0.14$ is the only value considered in \cite{Ulm} for the same $P_B$ constraint. Considering a set of crossover values rather than a single value provides more flexibility in designing error-correction codes. Our crossover probability range also allows us to use higher-rate codes than the codes for the range $[0.12, 0.14]$ since the maximum key rate $R_s^*$ of the fuzzy commitment scheme increases with decreasing $p_b$. The proposed transform-coding algorithm with the first quantizer method is thus beneficial for code design due to smaller crossover probability $p_b$. 

The maximum number of extracted bits, which corresponds to $N$ in (\ref{eq:Lpb}), for an $\displaystyle 8\!\times \!8$ RO array is $16$ bits for the \emph{1-out-of-8 masking} scheme \cite{ROPUFFirst}, $32$ bits for the \emph{non-overlapping RO pairs} \cite{ROPUFFirst}, and $64$ bits for the \emph{regression-based distillers} \cite{Yin2013improving}. Even if one assumes no errors, i.e., $R_s^*\!=\!1$, for these methods, their $S_{\text{max}}$ results are much smaller than the $S_{\text{max}}$ results of our algorithm, as shown in Fig.~\ref{fig:maxsecretkey}.

\begin{figure}[t]
	\centering
	\newlength\figureheight
	\newlength\figurewidth
	\setlength\figureheight{8.05cm}
	\setlength\figurewidth{15.1cm}
	\input{./maxsecretkeysize.tikz}
	\caption{The maximum key lengths $\displaystyle S_{\text{max}}$ for $\displaystyle 8\!\times \!8$ RO arrays.} 
	\label{fig:maxsecretkey}
\end{figure}
\subsection{Transform Complexity}
We measure the complexity of a transform in terms of the number of operations required to compute the transform and the hardware area required to implement it in a field-programmable gate array (FPGA). We are first interested in a computational-complexity comparison for RO arrays of sizes $ r\!=\!c\!=\!8$ and $r\!=\!c\!=\!16$, which are powers of 2, so that fast algorithms are available for the DCT, DWHT, and DHT. We then present an RO PUF hardware design for the transform with the minimum computational complexity. 

The computational complexity of the KLT for $r\!=\!c\!=\!N$ is $\displaystyle O(N^3)$, while it is $\displaystyle O(N^2\log_2 N)$ for the DCT and DWHT, and $\displaystyle O(N^2)$ for the DHT \cite{mySPbook}. There are efficient implementations of the DWHT without multiplications \cite{DWHTnomultiplication}. The DWHT is thus a good candidate for RO PUF designs for, e.g., internet of things (IoT) applications.

We now give a reference FPGA implementation for the DWHT without multiplications to illustrate that the hardware area occupied by the transform-coding algorithm is small and the processing time is significantly better than previous RO PUF designs.

\subsubsection{FPGA Implementation}
\begin{figure}[t]
	\centering
	\includegraphics[width=0.805\textwidth, height=0.785\textheight, keepaspectratio]{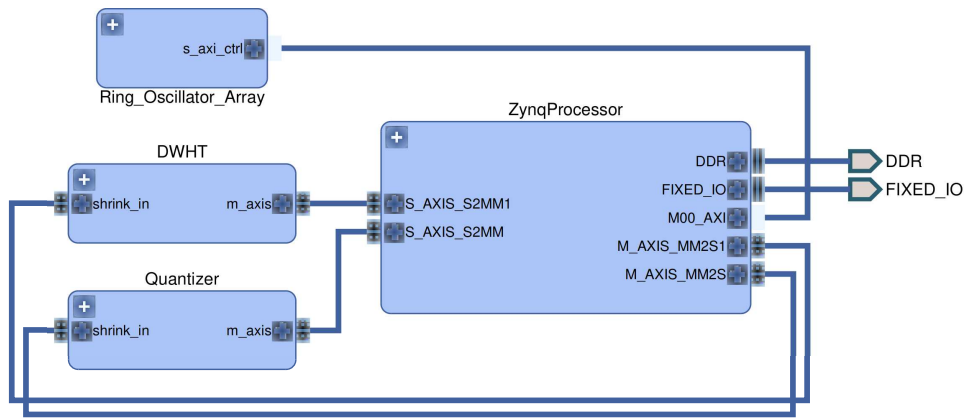}
	\caption{Hardware design overview.}
	\label{fig:hwOverview}
\end{figure}

We use a Xilinx ZC706 evaluation board with a Zynq-7000 XC7Z045 system-on-chip (SoC) to evaluate our DWHT design. A high level overview of the design is depicted in Fig.~\ref{fig:hwOverview}. The Zynq SoC consists of an FPGA part and an ARM Cortex-A9 dual-core processor, connected with memory-mapped AXI4 buses \cite{axi4}. The ARM processor is connected to three components: the RO array, DWHT, and quantizer. The RO array is connected via a bi-directional memory-mapped AXI bus, and the other components are connected via AXI streaming buses \cite{axiS}. We first measure RO outputs with counters, give the counter values as input to the DWHT, and then quantize the transform coefficients to assign bits. This is an implementation of the transform-coding algorithm given in Fig.~\ref{fig:postprocessing}.

We use a standard RO array of size $16\times16$. All ROs in a row are connected to a counter and ROs in the same row can be measured serially by using the counter. There is an additional counter that stops the counting operations after a specified time. For the FPGA we use, it is practically necessary to use at least five inverters for each RO since using three inverters results in oscillation frequencies of about 1GHz, which violates the timing constraints of the FPGA. Our RO designs with five inverters operate reliably and give oscillation frequencies in the range $[400, 500]$ MHz. Furthermore, we use 16-bit counters so that the minimum duration $T_{\text{min}}$ to have an overload in a counter is
\begin{align}
    T_{\text{min}} &= \frac{2^{16} - 1}{500\text{MHz}} = 131\mu\text{s}.\label{eq:Tmin}
\end{align}
We therefore count each RO output for a duration of $100\mu$s, which is less than $T_{\text{min}}$ to avoid overloads. This results in a total counting duration of $1.6$ms for all $16$ columns of the RO array, which is compared below with the previous RO PUF designs.

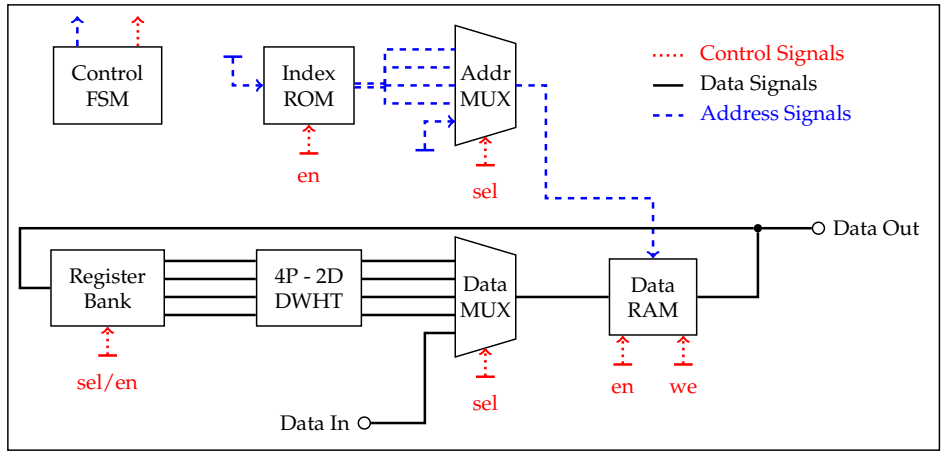
\begin{figure}
	\centering
	\resizebox{0.8\linewidth}{!}{
\begin{tikzpicture}[
    line width=0.7pt,
    ampersand replacement=\&,
    every text node part/.style={align=center}, 
    hwBB/.style={draw, rectangle, inner sep=3mm},
    conn point/.style={draw, fill, circle, inner sep=0.4mm},
    io port/.style={draw, circle, inner sep=0.7mm},
    ctrlSig/.style={red,   very thick, dotted},
    addrSig/.style={blue,  very thick, dashed},
    dataSig/.style={black, very thick},
]
    \tikzmath{
        \sigDist = 0.5;
    }

    \matrix[
        row sep=1.5cm,
        column sep=1.5cm,
    ] (hwMtx) {
        \node[hwBB] (fsm) {Control\\FSM};                     \&
        \node[hwBB]       (indexRom) {Index\\ROM};            \&
        \pic["Addr\\MUX"] (addrMux)  {mux=5/2.0/1.0/0.7/0.2}; \&
        \coordinate (legendPos);                              \\
        \node[hwBB]       (regBank) {Register\\Bank};        \&
        \node[hwBB]       (dwht)    {4P - 2D\\DWHT};          \&
        \pic["Data\\MUX", shift=(down:0.15)] (dataMux) {mux=5/2.0/1.0/0.7/0.2}; \&
        \node[hwBB, shift=(down:0.15)]       (dataRam) {Data\\RAM};             \\
    };
    \draw[ctrlSig] ($ (legendPos) + (0,  \sigDist) $) -- ++(\sigDist, 0) node [label={[name={lbl legT}] right:Control Signals}] {};
    \draw[dataSig] ($ (legendPos) + (0,  0)        $) -- ++(\sigDist, 0) node [label={right:Data Signals}] {};
    \draw[addrSig] ($ (legendPos) + (0, -\sigDist) $) -- ++(\sigDist, 0) node [label={right:Address Signals}] {};

    \draw[ctrlSig, ->] ($ (fsm.north) + ( \sigDist, 0) $) -- ++(0, \sigDist);
    \draw[addrSig, ->] ($ (fsm.north) + (-\sigDist, 0) $) -- ++(0, \sigDist);

    \draw[ctrlSig, |->] ($ (indexRom.south) + (0,         -\sigDist)$) node [label={below:en}] {}     -- ++(0, \sigDist);
    \draw[ctrlSig, |->] ($ (addrMux-sel)    + (0,         -\sigDist)$) node [label={below:sel}] {}    -- ++(0, \sigDist);
    \draw[ctrlSig, |->] ($ (regBank.south)  + (0,         -\sigDist)$) node [label={below:sel/en}] {} -- ++(0, \sigDist);
    \draw[ctrlSig, |->] ($ (dataMux-sel)    + (0,         -\sigDist)$) node [label={below:sel}] {}    -- ++(0, \sigDist);
    \draw[ctrlSig, |->] ($ (dataRam.south)  + (-\sigDist, -\sigDist)$) node [label={below:en}] {}     -- ++(0, \sigDist);
    \draw[ctrlSig, |->] ($ (dataRam.south)  + ( \sigDist, -\sigDist)$) node [label={below:we}] {}     -- ++(0, \sigDist);

    \draw[addrSig, |->] ($ (indexRom.west) + (-\sigDist, \sigDist)$)  |- (indexRom.west);
    \draw[addrSig, |->] ($ (addrMux-in4)   + (-\sigDist, -\sigDist)$) |- (addrMux-in4);

    \draw[addrSig, ->]    (addrMux-out) -- ++(\sigDist, 0) |- ($ (dataRam.north) + (0, 2*\sigDist) $) -| (dataRam.north);
    \draw[addrSig,double] (indexRom.east) -- ++(\sigDist, 0) coordinate (indexRom addrOut);
    \draw[addrSig] (indexRom addrOut) |- (addrMux-in0);
    \draw[addrSig] (indexRom addrOut) |- (addrMux-in1);
    \draw[addrSig] (indexRom addrOut) |- (addrMux-in2);
    \draw[addrSig] (indexRom addrOut) |- (addrMux-in3);

    \node[io port, label={[name={lbl dIn}] left:Data In}]    at ($ (dataMux-in4)        + (-3*\sigDist, -3*\sigDist) $) (dIn)     {};
    \node[conn point]                                   at ($ (dataRam.north east) + ( 2*\sigDist,  \sigDist) $)   (outConn) {};
    \node[io port, label={[name={lbl dOut}] right:Data Out}] at ($ (outConn)            + ( 2*\sigDist,  0) $)          (dOut)    {};

    \coordinate (regBank inLeft) at ($ (regBank.west) + (-\sigDist, 0) $);
    \draw[dataSig] (dIn) -| ($ (dataMux-in4) + (-\sigDist, 0) $) -- (dataMux-in4);
    \draw[dataSig] (regBank.west) -| (regBank inLeft) |- (outConn) -- (dOut);

    \draw[dataSig] (dataRam.east) -| (outConn);

    \foreach \i in {0,...,3} {
        \draw[dataSig] (regBank.east |- dataMux-in\i) -- (dwht.west |- dataMux-in\i);
        \draw[dataSig] (dwht.east |- dataMux-in\i) -- (dataMux-in\i);
    }
    \draw[dataSig] (dataMux-out) -- (dataRam.west);

    \node[draw, inner sep=2mm, fit=(hwMtx) (regBank inLeft) (lbl dIn) (lbl dOut) (lbl legT)] {};
\end{tikzpicture}
	}
	\caption{Building blocks for the DWHT implementation.}\label{fig:hwDwht}
\end{figure}

We next implement an extended version of the algorithm, proposed for an $8\times 8$ array, in \cite{DWHTnomultiplication} to calculate the two-dimensional (2D) $16\times 16$ DWHT without multiplications. The main block we use is the 4-point (4P)-2D DWHT \cite{DWHTnomultiplication} that takes four inputs $[x_0, x_1, x_2, x_3]$ and calculates
\begin{align}
\begin{bmatrix} y_0& y_1\\y_2&y_3 \end{bmatrix} = 
\frac{1}{2} 
\begin{bmatrix}
    x_0+x_1+x_2+x_3 &  x_0-x_1+x_2-x_3\\
    x_0+x_1-x_2-x_3 & x_0-x_1-x_2+x_3 
\end{bmatrix}.\label{eq:4P2DDWHT}
\end{align}
We successively apply the 4P-2D DWHT to the $16\times 16$ RO array according to an extension of the input-selection algorithm proposed in \cite{DWHTnomultiplication}. We implement a finite state machine (FSM) to control the input and output AXI streaming interfaces as well as the input-selection algorithm. The building blocks of our DWHT implementation is depicted in Fig.~\ref{fig:hwDwht}, which includes
\begin{itemize}
    \item a data random access memory (RAM) to store all array elements,
    \item a 32-bit index read-only memory (ROM), where each word stores four 8-bit array-element addresses,
    \item a multiplexer (MUX) to select the RAM address to be accessed,
    \item a second MUX to select the ROM input,
    \item a register for each input to convey different RAM words to different ports.
\end{itemize} 

We first store all RO outputs in the data RAM. Then, the first word of the index ROM is fetched. This word holds the addresses of four array elements to be loaded. These array elements are passed to the 4P-2D DWHT's input registers by selecting the corresponding port in the address MUX and register bank. After evaluating the 4P-2D DWHT, the new array elements $[y_0,y_1,y_2,y_3]$ are written back to the locations from where the inputs $[x_0, x_1, x_2, x_3]$ were fetched. The FSM performs the same steps for all remaining ROM words and conveys the 2D DWHT coefficients to the AXI output port. 

The addition and subtraction operations on four numbers in each 4P-2D DWHT evaluation requires at most two additional bits, while the subsequent bit shift to implement the division by $2$ in (\ref{eq:4P2DDWHT}) removes one bit. Since the 4P-2D DWHT is applied in total four times to each RAM location, the transform requires $20$-bit operations and storage in order to process the $16$-bit signed numbers used for counter values.

The quantizer contains AXI stream ports, an FSM, and one ROM. The ROM holds $2^{K_i}-1$ quantization boundaries for the $i$-th transform coefficient. We remark that the histogram equalization step in Fig.~\ref{fig:postprocessing} is useful when the number of bits $K_i$ extracted are large, but we choose $K_i=K=1$ for all used transform coefficients, which is illustrated in combination with an error-correction code design in Section~\ref{subsec:codedesignforquant2}. Therefore, we do not apply the histogram equalization step for this case, so the ROM contains $255$ words and is of size $638$ Bytes ($\geq 255*20$ bits) in total. The FSM compares the quantizer input with the corresponding quantization boundary to assign a bit $1$ for transform-coefficient values greater than the quantization boundary, and the bit $0$ otherwise. The assigned bits are then conveyed to the output port.

\subsubsection{Hardware Design Comparisons}
We now compare our results with another RO PUF hardware design given in \cite{Pufky} in terms of the hardware area and processing times. The number of LUTs, registers, and MUXs used in \cite{Pufky} are not available. However, our results can be compared with their slice-count and processing-delay results since the FPGA (Spartan-6) used in \cite{Pufky} also has 4 LUTs, 8 registers and 3 MUXes in each slice, the same as the FPGA used in this work. In addition, the quantizer and DWHT clock rate is $54$MHz, as in \cite{Pufky}. There are alternative RO PUF designs in \cite{ref1,ref2}, but their secret-key lengths are smaller than $128$ bits, which makes a comparison with our scheme difficult. Therefore, we list in Table~\ref{tab:hwResults} the hardware area occupied by individual components of our RO PUF design and by the RO PUF design of \cite{Pufky}.

Table~\ref{tab:hwResults} illustrates that the RO array causes the highest hardware cost and uses approximately $82\%$ of all occupied LUTs, $62\%$ of registers, and $86\%$ of slices. We do not include the area for RAMs and ROMs, because we use Block RAM slices that are available in the FPGA. However, we include the control logic area required to control the Block RAM slices. Our DWHT-based design occupies an approximately $11\%$ smaller RO PUF hardware area than the RO PUF design proposed in \cite{Pufky} in terms of the number of slices used. This result can be improved if we re-use the same area for different ROs, which might increase correlations in the RO outputs. In addition, the DWHT and quantizer constitute approximately $14\%$ of the total slice count of our RO PUF design. These results illustrate that the transform-coding approach occupies a small hardware area.

The total counter duration of $1.6$ms is a result of the calculation given in (\ref{eq:Tmin}) to avoid overloads in the counters, and the choice of this value depends mainly on the number of inverters used for each RO and counter bit width. The overall processing time of the proposed design is approximately $1.68$ms, which is significantly better than the processing delay of the RO PUF design in \cite{Pufky}.

\begin{table}[t]
	\centering
	\caption{Hardware area and processing delays for RO PUF designs.}
	\begin{tabular}{ |c|c|c|c|c|c|c| }
		\hline
		\rowcolor{blue!25}
		Blocks &  
		\textbf{LUTs} &
		\textbf{Registers} &
		\textbf{MUXes} &
		\textbf{RAM\&ROM} $\left[\text{Byte}\right]$ &
		\textbf{Slices} &
		\textbf{Duration} $\left[\mu s\right]$\\
		\hline
		Proposed-ROs       & 1632 & 397  & 65   & 0    & 729 & 1600\\
		\hline
		Proposed-DWHT      & 326  & 200  & 0    & 1664 & 99  & 66\\
		\hline
		Proposed-Quantizer & 43   & 39   & 0    & 638  & 21  & 14\\
		\hlinewd{1.2pt}
		Proposed (ROPUF)   & 2001 & 636  & 65   & 2302 & 849 & 1680\\
		\hline
		PUFKY (ROPUF) \cite{Pufky} 
		& n.a. & n.a. & n.a. & n.a. & 952 & 4611\\
		\hline
	\end{tabular}\label{tab:hwResults}
\end{table}

\subsection{Uniqueness and Security}\label{subsec:uniqueness}
The bit sequence extracted from a physical identifier should consist of uniformly distributed bits so that the rate region $\mathcal{R}$ in (\ref{eq:ls0}) is valid. A common measure, called \textit{uniqueness}, for checking randomness of a bit sequence is the average fractional Hamming distance between the bit sequences extracted from different RO PUFs \cite{bizimpaper}. We obtain similar uniqueness results for all transforms, where the mean Hamming distance is $0.500$ and Hamming distance variance is approximately $\displaystyle 7\!\times \!10^{-4}$. All transforms thus provide close to optimal uniqueness results due to their high decorrelation efficiencies and equipartitioned quantization intervals. These results are significantly better than the results $0.462$ \cite{ROPUFFirst} and $0.473$ \cite{ROLarge}.  

The National Institute of Standards and Technology (NIST) provides a set of randomness tests that check whether a bit sequence can be differentiated from a uniformly random bit sequence \cite{NIST}. We apply these tests to evaluate the randomness of the generated sequences. We observe that the bit sequences generated from ROs in the dataset \cite{ROLarge} with the DWHT pass most of the applicable tests for short lengths for both reliability metrics, which is considered to be an acceptable result \cite{NIST}. We also conclude that the KLT performs the best due to its optimal decorrelation performance. One can apply a thresholding approach such that the reliable transform coefficients from which the bits are extracted do not have high correlations, which further improves the security performance \cite{bizimtemperature}.

\section{Privacy and Secrecy Analysis of Proposed Error-correction Codes}\label{sec:correction}
Suppose that extracted bit sequences are uniformly distributed so that the secrecy leakage is zero. We propose different codes for the transform-coding algorithm according to the two proposed reliability metrics.

\subsection{Codes for the Quantizer Design with Fixed Measurement Channels}\label{subsec:CodesforQuant1}
For the first quantizer method given in Section~\ref{subsec:quant1}, fix an average crossover probability $\displaystyle p_b\!=\!0.06$ to obtain the highest maximum secret-key length, as shown in Fig.~\ref{fig:maxsecretkey}. We illustrate that there are efficient error-correction codes for the fuzzy commitment scheme with $P_B\!\leq\!\displaystyle 10^{-9}$ and a small privacy-leakage rate. Recall that the code dimension has to be at least $128$ bits, a requirement of the AES, so the block length is in the short block-length regime for error-correction codes with high rates and $k\! =\! 128$. We expect a rate loss in our code designs due to the small block-error probability constraint and short block length. One needs finite-length bounds for the fuzzy commitment scheme, which are not available in the literature. We thus compare the performance of our codes with the region $\mathcal{R}$ given in (\ref{eq:ls0}). The basic approach to design codes for small block-error probabilities and reasonable decoding complexity is to use concatenated codes. Since the hardware complexity of a code design should be small for IoT applications, we minimize also the field sizes of the codes. 

\begin{Remark}
	\normalfont It would be natural to use iterative decoders in combination with high-performance codes like low density parity check (LDPC) and turbo codes. However, hardware complexity might increase and it is a difficult task to simulate these codes for $P_B\!\leq\!10^{-9}$. We thus use concatenated algebraic codes so that we can find analytical bounds on $P_B$ without simulations for the outer code. 
\end{Remark}

The first construction uses a Reed-Muller (RM) code $\mathcal{C}(32,6,16)$ as the inner code and a Reed-Solomon (RS) code $\mathcal{C}(28,22,7)$ that operates with symbols from the Galois field $\mathbb{F}_{2^6}$ as the outer code of a concatenated code. Every symbol of the RS code can be represented by 6 bits and the code takes $22$ symbols as input, which corresponds to $132$ input bits that is greater than $128$ bits. The majority logic decoder (MLD) of the inner RM code transforms the BSC with crossover probability $\displaystyle p_b\!=\!0.06$ into a channel with errors and erasures by declaring an \emph{erasure} if there are two codewords with equal distances to a received vector and makes an \emph{error} if a wrong codeword is selected. Simulation results show that the erasure probability after the MLD of the inner code is about $\displaystyle 6.57\!\times\!10^{-5}$ and the error probability is about $\displaystyle 4.54\!\times\!10^{-6}$. The BMDD for the outer code correctly reconstructs the codeword if $2\cdot e\! +\! \nu \!<\! d$, where $e$ is the number of errors and $\nu$ is the number of erasures in the received vector \cite{ECC}. The block-error probability after decoding the outer RS code is approximately $P_B\!\approx\!1.37\!\times\!10^{-11}$. The key and leakage rates of this code are $R_s\!=\! 0.1473$ and $R_l\!=\!0.8527$ bits/source-bit, respectively.

An alternative concatenated code is a binary extended Bose-Chaudhuri-Hocquenghem (BCH) code $\displaystyle\mathcal{C} (256,132,36)$ as the outer code and a repetition code $\displaystyle \mathcal{C}(3,1,3)$ as the inner code. The maximum-likelihood decoder for the inner code transforms the BSC with crossover probability $p_b\!=\!0.06$ into a BSC with $p_b\!=\!0.0104$ so that the BMDD for the outer BCH code results in $P_B\!=\!\displaystyle 3.48\!\times\!10^{-10}$. The key-leakage rate pair $\left(R_s,R_l\right)$ for this code is $\left(0.1719,0.8281\right)$ bits/source-bit, which gives better rates than the RM+RS concatenation above and the best generalized-concatenated-code (GCC) design with the fuzzy commitment scheme in \cite{Ulm} with the key-leakage rate pair $\left(0.1260,0.8740\right)$ bits/source-bit, which is shown to be better than the previous results in \cite{Pufky}. The significant improvement in the rates with a low-complexity code is due to the decrease in $p_b$ by using our transform-coding algorithm. 

The fuzzy commitment scheme can asymptotically achieve the maximum secret-key rate $R_s^*=0.6726$ bits/source-bit and corresponding minimum privacy-leakage rate $R_l^*=0.3274$ bits/source-bit for a BSC$(p_b\!=\!0.06)$. Better key-leakage rate pairs are thus possible, e.g., by using GCCs or by improving the decoder for the outer code. However, these constructions would result in increased hardware complexity, which is not desired for IoT applications.

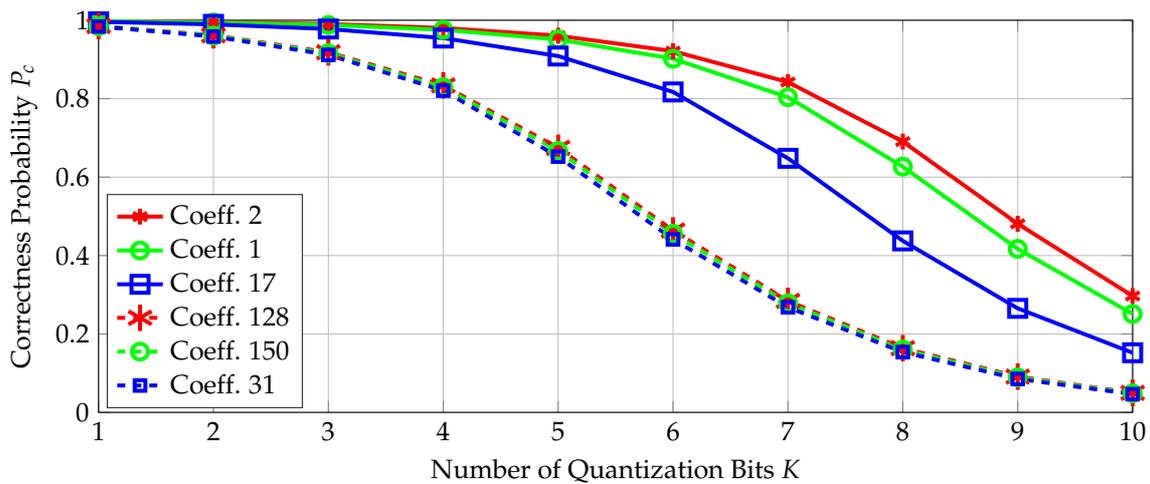
\begin{figure}[t]
\centering
\setlength\figureheight{5.2cm}
\setlength\figurewidth{14.3cm}
%
%
\begin{tikzpicture}

\begin{axis}[%
width=0.951\figurewidth,
height=\figureheight,
at={(0\figurewidth,0\figureheight)},
scale only axis,
xmin=1,
xmax=10,
xlabel={Number of Quantization Bits $K$},
xmajorgrids,
ymin=0,
ymax=1,
ylabel={Correctness Probability $P_c$},
ymajorgrids,
axis background/.style={fill=white},
legend style={at={(0.0079,0.0107)},anchor=south west,legend cell align=left,align=left,draw=white!15!black}
]
\addplot [color=red,solid,line width=1.5pt,mark size=3.0pt,mark=asterisk,mark options={solid}]
  table[row sep=crcr]{%
1	0.998257301311316\\
2	0.995481003281785\\
3	0.990369641601798\\
4	0.98040809110094\\
5	0.960634304705393\\
6	0.92117005238719\\
7	0.842292962766348\\
8	0.690643212598228\\
9	0.481072154279916\\
10	0.297004683385764\\
11	0.17184382736417\\
12	0.0959793040700199\\
};
\addlegendentry{Coeff. 2};

\addplot [color=green,solid,line width=1.5pt,mark size=3.0pt,mark=o,mark options={solid}]
  table[row sep=crcr]{%
1	0.997829854852876\\
2	0.994372582050197\\
3	0.988007494814884\\
4	0.975602554847855\\
5	0.95097861539375\\
6	0.901834495690211\\
7	0.803718836699901\\
8	0.626612080739878\\
9	0.417072421387783\\
10	0.251075314644554\\
11	0.143338414886165\\
12	0.0794353605101738\\
};
\addlegendentry{Coeff. 1};

\addplot [color=blue,solid,line width=1.5pt,mark size=3.2pt,mark=square,mark options={solid}]
  table[row sep=crcr]{%
1	0.995957945354216\\
2	0.989518408888429\\
3	0.97766272562568\\
4	0.954557152546349\\
5	0.908692362891015\\
6	0.817204744400462\\
7	0.647964289458797\\
8	0.43737401312202\\
9	0.26530922001697\\
10	0.152081710529126\\
11	0.0844834747559681\\
12	0.0460313603210151\\
};
\addlegendentry{Coeff. 17};

\addplot [color=red,dashed,line width=1.5pt,mark size=5pt,mark=asterisk,mark options={solid}]
  table[row sep=crcr]{%
1	0.985222877647597\\
2	0.961673818771534\\
3	0.918315977543005\\
4	0.833830216915307\\
5	0.674630893567695\\
6	0.463852936793098\\
7	0.284268575193586\\
8	0.163836393340128\\
9	0.0913018064965708\\
10	0.0498483977069998\\
11	0.0268445209066746\\
12	0.0143139766736807\\
};
\addlegendentry{Coeff. 128};

\addplot [color=green,dashed,line width=1.5pt,mark size=3pt,mark=o,mark options={solid}]
  table[row sep=crcr]{%
1	0.984810457218178\\
2	0.960603711151823\\
3	0.916034816384811\\
4	0.829195710003641\\
5	0.66670380695252\\
6	0.455725042335783\\
7	0.278372640482159\\
8	0.160160315818596\\
9	0.0891633999024296\\
10	0.048649271931228\\
11	0.0261868464961171\\
12	0.013958574261824\\
};
\addlegendentry{Coeff. 150};

\addplot [color=blue,dashed,line width=1.5pt,mark size=2pt,mark=square,mark options={solid}]
  table[row sep=crcr]{%
1	0.984039637730627\\
2	0.95860356541513\\
3	0.911770979982468\\
4	0.820540284873643\\
5	0.652232492013706\\
6	0.441272707302553\\
7	0.268015323957184\\
8	0.153736674739541\\
9	0.0854365302130378\\
10	0.0465625576457378\\
11	0.0250434748870105\\
12	0.013341123560002\\
};
\addlegendentry{Coeff. 31};

\end{axis}
\end{tikzpicture}%
\caption{The correctness probabilities for transform coefficients.} 
\label{fig:correctnessprob}
\end{figure}

\subsection{Codes for the Quantizer Design with Fixed Number of Errors}\label{subsec:codedesignforquant2}
We now select a channel code according to Section~\ref{subsec:quant2} to store a secret key of length 128 bits. The correctness probabilities defined in (\ref{eq:correctness}) for the transform coefficients $T$ with the three highest and three smallest probabilities are plotted in Fig.~\ref{fig:correctnessprob}. The indices of the $16\times 16$ transform coefficients follow the order in the dataset \cite{ROLarge}, where the coefficient index at the first row and first column is $1$, and it increases columnwise up to $16$ so that the second row starts with the index $17$, the third row with the index $33$, etc. The most reliable transform coefficients are the low-frequency coefficients, which are in our case at the upper-left corner of the 2D transform-coefficient array with indices such as $1,2,3,17,18,19,33,34,35$. The low-frequency transform coefficients therefore have the highest SNRs for the source and noise statistics obtained from the RO dataset in \cite{ROLarge}. The least reliable coefficients are observed to be spatially away from the transform coefficients at the upper-left or lower-right corners of the 2D transform-coefficient array. These results indicate that the \emph{SNR-packing efficiency}, which can be defined similarly as the energy-packing efficiency, of a transform follows a more complicated scan order than the classic zig-zag scan order used for the energy-packing efficiency metric \cite{Zigzag}. Observe from Fig.~\ref{fig:correctnessprob} that increasing the number of extracted bits decreases the correctness probability for all coefficients since the quantization boundaries get closer so that errors due to noise become more likely, i.e., the probability $P_{c}(K)$ defined in (\ref{eq:correctness}) decreases with increasing $K$.

We fix the maximum number $\displaystyle C_{\text{max}}$ of transform coefficients $T$ allowed to be in error and calculate the correctness threshold $\displaystyle \xbar{P}_c(C_{\text{max}})$ using (\ref{eq:threshold}), the total number $\displaystyle N(C_{\text{max}})$ of extracted bits using (\ref{eq:totalbits}), and the number $\displaystyle e(C_{\text{max}})$ of errors the block code should be able to correct using (\ref{eq:minimume}). We observe that if $\displaystyle C_{\text{max}}\!\leq\!10$, $\xbar{P}_c(C_{\text{max}})$ is so large that $\displaystyle P_{c,i}(K\!=\!1)\!\leq\!\xbar{P}_c(C_{\text{max}})$ for all $i=2,\ldots,L$. If $\displaystyle 11\!\leq\!C_{\text{max}}\!\leq\!15$, $\displaystyle N(C_{\text{max}})$ is less than the required code dimension of 128 bits. Increasing $\displaystyle C_{\text{max}}$ results in a smaller correctness threshold $\displaystyle \xbar{P}_c(C_{\text{max}})$ so that the maximum of the number $\displaystyle K_{\text{max}}(C_{\text{max}})\!=\!K^{\prime}_1(C_{\text{max}})$ of bits extracted among the $L-1$ used coefficients increases. This approach can increase hardware complexity. We thus do not consider the cases where $\displaystyle C_{\text{max}}\!>\!20$. Table~\ref{tab:myresults} shows $\displaystyle \xbar{P}_c(C_{\text{max}})$, $\displaystyle N(C_{\text{max}})$, and $\displaystyle e(C_{\text{max}})$ for the remaining range of $\displaystyle C_{\text{max}}$ values, which are used for channel-code selection.

Consider again binary (extended) BCH and RS codes, which have good minimum-distance properties. An exhaustive search does not provide a code with dimension of at least 128 bits and with parameters satisfying any of the  $\displaystyle (N(C_{\text{max}}),\,e(C_{\text{max}}))$ pairs in Table~\ref{tab:myresults}. However, the correctness threshold analysis leading to Table~\ref{tab:myresults} is conservative. We therefore choose a BCH code with parameters as close as possible to a $\displaystyle (N(C_{\text{max}}),\,e(C_{\text{max}}))$ pair and then prove that even if the number $\displaystyle e_{\text{BCH}}$ of errors the chosen BCH code can correct is less than $\displaystyle e(C_{\text{max}})$, the block-error probability constraint is satisfied. Consider therefore the BCH code with the block length $255$, code dimension $131$, and a capability of correcting all error patterns with $\displaystyle e_{\text{BCH}}=18$ or less errors.

We now show that the proposed code satisfies the block-error probability constraint. First, we impose the condition that exactly one bit is extracted from each coefficient, i.e., $K_{i}\!=\!1$ for all $i\!=\!2,3,\ldots,L$, so that in total $N\!=\!L-1\!=\!255$ bits are obtained. Note that this results in independent bit errors $E_i$. It follows from this condition that the chosen block code should be able to correct all error patterns with up to $e\!=\!20$ bit errors rather than $e(20)\!=\!25$ bit errors, which is still greater than the error-correction capability $\displaystyle e_{\text{BCH}}=18$ of the considered BCH code.

The block error probability $P_B$ for the BCH code $\mathcal{C}(255,131,37)$ with a BMDD corresponds to the probability of having more than $18$ errors in the codeword, i.e.,
\begin{align}
 P_B = \sum_{j=19}^{255}\Bigg[\sum_{A\in\mathcal{F}_j}\prod_{i\in A}(1-P_{c,i})\,\bigcdot\prod_{i\in A^{c}}P_{c,i} \Bigg] \label{eq:blockerrorforbch}
\end{align}
where $P_{c,i}$ is the correctness probability of the $i$-th transform coefficient $\widehat{T}_i$ defined in (\ref{eq:correctness}) for $i\!=\!2,3,\ldots,256$, $\displaystyle \mathcal{F}_j$ is the set of all size-$j$ subsets of the set $\displaystyle\{2,3,\ldots,256\}$, and $A^{c}$ denotes the complement of the set $A$. The correctness probabilities $P_{c,i}$ are different and they represent probabilities of independent events due to the independence assumption for the transform coefficients. 

\begin{table}[t]
\centering
\caption{Code-parameter constraints.}
\begin{tabular}{ |c|c|c|c|c|c| }
\hline
\rowcolor{blue!25}
$\displaystyle \mathbf{C_{\text{\textbf{max}}}}$ & $\mathbf{16}$ & $\mathbf{17}$& $\mathbf{18}$ & $\mathbf{19}$ & $\mathbf{20}$\\
\hline
$\displaystyle \bar{P}_c$ &  $0.9902$
 &  $0.9889$  &   $0.9875$ &  $0.9860$ &  $0.9844$\\
\hline
$\displaystyle K_{\text{max}}$ &  $3$ &  $3$ &  $3$ & $3$ & $3$\\
\hline
$\displaystyle N$ & $144$ & $224$ &  $250$ & $255$ & $259$\\
\hline
$\displaystyle e$ &  $18$ &  $20$&  $21$ & $23$ & $25$\\
\hline
\end{tabular}\label{tab:myresults}
\end{table}

One needs to consider $ \sum_{j=0}^{18}{255\choose j}\approx 1.90\!\times\!10^{27}$ different cases to calculate (\ref{eq:blockerrorforbch}), which is not practical. We thus use the discrete Fourier transform - characteristic function (DFT-CF) method \cite{DFTCF} to calculate the block-error probability and obtain the result $P_B\!\approx\!1.26\!\times\!10^{-11}\!<\!10^{-9}$. The block-error probability constraint is thus satisfied by using the BCH code $\mathcal{C}(255,131,37)$ with a BMDD although the conservative analysis suggests that it would not be satisfied.

We now compare the BCH code $\mathcal{C}(255,131,37)$ with previous codes proposed for binding keys to physical identifiers with the fuzzy commitment scheme and a secret-key length of $128$ bits such that $P_B\!\leq\!10^{-9}$ is satisfied. The (secret-key, privacy-leakage) rate pair for this proposed code is $(R_s,R_l)=(\frac{131}{255},1\!-\!\frac{131}{255})\approx(0.514,\,0.486)$ bits/source-bit. This pair is significantly better than our previous results in Section~\ref{subsec:CodesforQuant1} proposed for a BSC$(p_b\!=\!0.06)$. The main reason for obtaining a better (secret-key, privacy-leakage) rate pair is that the quantizer in Section~\ref{subsec:quant2} allows us to exploit higher identifier-output reliability by decreasing the number of bits extracted from each transform coefficient. 

\begin{figure}[t] 
	\centering
	\setlength\figureheight{5.75cm}
	\setlength\figurewidth{14.4cm}
	\input{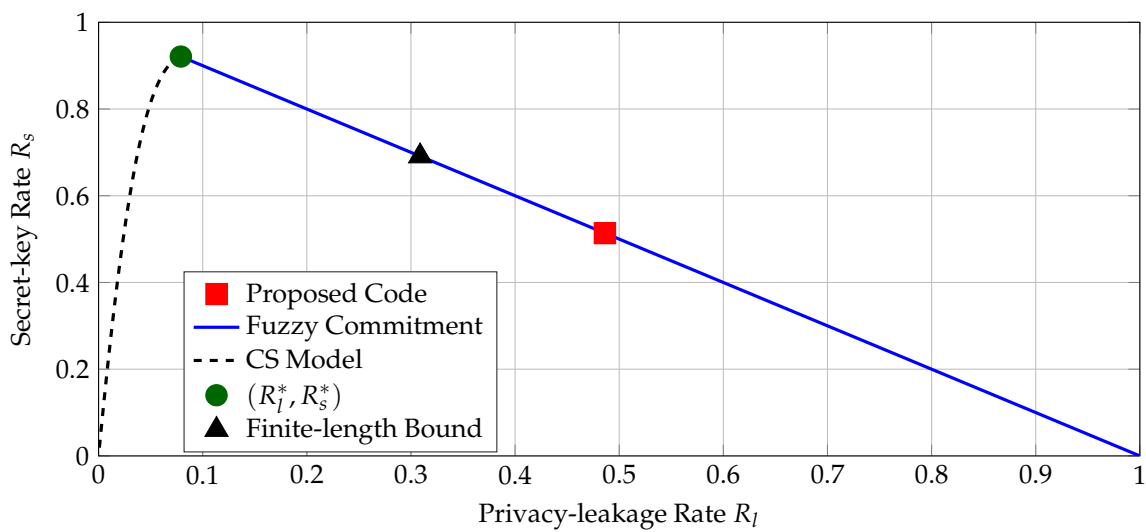}
	\caption{The operation point of the proposed BCH code $\mathcal{C}(255,131,37)$, regions of achievable rate pairs according to (\ref{eq:ls0}) and (\ref{eq:chosensecret}), the maximum secret-key rate point, and a finite-length bound for $N=255$ bits, $P_B=10^{-9}$, and BSC$(0.0097)$.} 
	\label{fig:ratecomparison}
\end{figure}

We compare the secret-key and privacy-leakage rates of the BCH code $\mathcal{C}(255,131,37)$ with the region of all achievable rate pairs for the CS model and the fuzzy commitment scheme for a BSC $P_{Y|X}$ with crossover probability $p_b\!=\! 1-\frac{1}{L-1}\sum_{i=2}^LP_{c,i}(K_i\!=\!1)\!\approx\!0.0097$, i.e., the probability of being in error averaged over all used transform coefficients with the quantizer in Section~\ref{subsec:quant2}. We compute the boundary points of the region $\mathcal{R}_{\text{cs}}$ by using Mrs. Gerber's lemma \cite{WZ}, which gives the optimal auxiliary random variable $U$ in (\ref{eq:chosensecret}) when $P_{Y|X}$ is a BSC. We plot the regions of all rate pairs achievable with the fuzzy commitment scheme and CS model, the maximum secret-key rate point, the (secret-key, privacy-leakage) rate pair of the proposed code, and a finite-length bound \cite{Polyanskiy} for the block length of $N=255$ bits and $P_B\!=\!10^{-9}$ in Fig.~\ref{fig:ratecomparison}.

The maximum secret-key rate is $R_s^*\!\approx\!0.922$ bits/source-bit with a corresponding minimum privacy-leakage rate of $R_l^*\!\approx\!0.079$ bits/source-bit. There is a gap between the secret-key rate of the proposed code and the only operation point where the fuzzy commitment scheme is optimal. Part of this rate loss can be explained by the short block length of the code and the small block-error probability constraint. The finite-length bound given in \cite[Theorem 52]{Polyanskiy} establishes that the rate pair $(R_s,R_l)\!=\!(0.691,0.309)$ bits/source-bit is achievable by using the fuzzy commitment scheme, as depicted in Fig.~\ref{fig:ratecomparison}. One can therefore further improve the rate pairs by using better codes and decoders with higher hardware complexity, but this may not be possible for IoT applications. Fig.~\ref{fig:ratecomparison} also illustrates that there exist other code constructions, e.g., the WZ-coding construction in \cite{bizimWZ}, that reduce the privacy-leakage rate for a fixed secret-key rate.

\section{Conclusion}\label{sec:conclusion}
The reliability, uniqueness, security, computational-complexity, and key-length performance of various transforms was compared to select the best transforms for reliable secret-key binding for RO PUFs by using the fuzzy commitment scheme. The DWHT and DHT perform best in terms of computational-complexity, maximum key length, and reliability. All transforms give close to optimal uniqueness and good security results. A reference hardware design with the DWHT showed that the hardware area required by the transform-coding approach is small and less than required by the existing RO PUF designs. Low-complexity concatenated codes with high secret-key and small privacy-leakage rates, which are better than previous results, are proposed for a realistic block-error probability of $10^{-9}$. 

We further improved the transform-coding algorithm applied to physical identifiers by designing quantizers with reliability guarantees. This alternative quantizer converts the block-error probability constraint $P_B\!\leq\!10^{-9}$ into a constraint on the number of transform coefficients allowed to be in error. We proposed a BCH code $\mathcal{C}(255,131,37)$ with a higher code rate than our previously proposed codes. Comparisons with the region of all achievable (secret-key, privacy-leakage) rate pairs for the fuzzy commitment scheme show that there is still a gap between the optimal rate pairs and the proposed code. This gap can be closed by using other channel codes and decoders at the cost of higher hardware complexity or by designing codes for other CS model constructions. In future work, we will apply an extension of water-filling techniques to the transform-coefficients in order to improve the reliability and security performance.


\vspace{6pt} 

\acknowledgments{O. G\"unl\"u thanks Anes Belkacem and Bernhard C. Geiger for their contributions to one of the conference papers used in this work. O. G\"unl\"u was supported by the German Research Foundation (DFG) through the project HoliPUF under the grant KR3517/6-1. V. Sidorenko is on leave from the Institute for Information Transmission Problems, Russian Academy of Science. G. Kramer was supported by an Alexander von Humboldt Professorship endowed by the German Federal Ministry of Education and Research.}
%

\reftitle{References}
\bibliographystyle{unsrt}

\begin{thebibliography}{-------}

\bibitem[G{\"u}nl{\"u} \em{et~al.}(2016)G{\"u}nl{\"u}, \.{I}\c{s}can,
Sidorenko, and Kramer]{ourGlobalSIP}
G{\"u}nl{\"u}, O.; \.{I}\c{s}can, O.; Sidorenko, V.; Kramer, G.
\newblock Reliable secret-key binding for physical unclonable functions with
transform coding.
\newblock  IEEE Global Conf. Sign. and Inf. Process.; Greater Washington, {DC}, Dec. 2016; pp. 986--991.

\bibitem[G{\"u}nl{\"u} \em{et~al.}(2017)G{\"u}nl{\"u}, Belkacem, and
Geiger]{bizimICC}
G{\"u}nl{\"u}, O.; Belkacem, A.; Geiger, B.C.
\newblock Secret-key binding to physical identifiers with reliability
guarantees.
\newblock  {IEEE} Int. Conf. Commun.; Paris, France, May 2017; pp. 1--6.

\bibitem[Suh \em{et~al.}(2003)Suh, Clarke, Gassend, Dijk, and
Devadas]{Suhaegis}
Suh, G.E.; Clarke, D.; Gassend, B.; Dijk, M.V.; Devadas, S.
\newblock {AEGIS}: Architecture for tamper-evident and tamper-resistant
processing.
\newblock  {ACM} 17th Annu. Int. Conf. Supercomputing; New York, {NY}, June 2003; pp. 160--171.

\bibitem[Pappu(2001)]{PappuThesis}
Pappu, R.
\newblock Physical one-way functions.
\newblock PhD thesis, M.I.T., Cambridge, MA, 2001.

\bibitem[B{\"o}hm and Hofer(2013)]{pufintheory}
B{\"o}hm, C.; Hofer, M.
\newblock {\em Physical unclonable functions in theory and practice}; Springer:
New York, NY,  2013.

\bibitem[Guajardo \em{et~al.}(2007)Guajardo, Kumar, Schrijen, and
Tuyls]{SRAMPUFFirst}
Guajardo, J.; Kumar, S.S.; Schrijen, G.J.; Tuyls, P.
\newblock {FPGA} intrinsic {PUF}s and their use for {IP} protection. In {\em
	Int. Workshop Cryptographic Hardware and Embedded Systems}; Paillier, P.;
Verbauwhede, I., Eds.; Berlin Heidelberg, Germany: Springer-Verlag,  2007;
pp. 63--80.

\bibitem[Suh and Devadas(2007)]{ROPUFFirst}
Suh, G.E.; Devadas, S.
\newblock Physical unclonable functions for device authentication and secret
key generation.
\newblock  {ACM/IEEE} Design Automation Conf.; San Diego, {CA}, June 2007; pp. 29--14.

\bibitem[Gassend(2003)]{GassendThesis}
Gassend, B.
\newblock Physical random functions.
\newblock Master's thesis, M.I.T., Cambridge, MA,  2003.

\bibitem[Dodis \em{et~al.}(2008)Dodis, Ostrovsky, Reyzin, and
Smith]{FuzzyBasic}
Dodis, Y.; Ostrovsky, R.; Reyzin, L.; Smith, A.
\newblock Fuzzy extractors: How to generate strong keys from biometrics and
other noisy data.
\newblock {\em Soc. Industrial Appl. Math. J. Comp.} Mar. {\bf 2008}, {\em
	38},~97--139.

\bibitem[Juels and Wattenberg(1999)]{juelsfuzzy}
Juels, A.; Wattenberg, M.
\newblock A fuzzy commitment scheme.
\newblock  {ACM} Conf. Comp. and Commun. Security; New York, {NY}, Nov. 1999; pp. 28--36.

\bibitem[G{\"u}nl{\"u} \em{et~al.}(2017)G{\"u}nl{\"u}, \.{I}\c{s}can,
Sidorenko, and Kramer]{bizimWZ}
G{\"u}nl{\"u}, O.; \.{I}\c{s}can, O.; Sidorenko, V.; Kramer, G.
\newblock Wyner-Ziv coding for physical unclonable functions and biometric
secrecy systems.
\newblock Sep. 2017, [Online]. Available: arxiv.org/abs/1709.00275.

\bibitem[Maes(2013)]{RoelMaesbook}
Maes, R.
\newblock {\em Physically unclonable functions}; Berlin-Heidelberg, Germany:
Springer-Verlag,  2013.

\bibitem[Ignatenko and Willems(2009)]{IgnatenkoTrans}
Ignatenko, T.; Willems, F.
\newblock Biometric systems: Privacy and secrecy aspects.
\newblock {\em IEEE Trans. Inf. Forensics and Sec.} Dec. {\bf 2009}, {\em
	4},~956--973.

\bibitem[Lai \em{et~al.}(2008)Lai, Ho, and Poor]{Lai10}
Lai, L.; Ho, S.W.; Poor, H.V.
\newblock Privacy-security trade-offs in biometric security systems - {P}art {I}: {S}ingle use case.
\newblock  {\em IEEE Trans. Inf. Forensics and Sec.} Mar. {\bf 2011}, {\em
	6},~122--139.

\bibitem[G{\"u}nl{\"u} and Kramer(2016)]{bizimproofarxiv}
G{\"u}nl{\"u}, O.; Kramer, G.
\newblock Privacy, secrecy, and storage with noisy identifiers.
\newblock Jan. 2016, [Online]. Available: arxiv.org/abs/1601.06756.

\bibitem[G{\"u}nl{\"u}(2013)]{benimthesis}
G{\"u}nl{\"u}, O.
\newblock Design and analysis of discrete cosine transform based ring
oscillator physical unclonable functions.
\newblock Master's thesis, Techn. Univ. Munich, Munich, Germany,  2013.

\bibitem[G{\"u}nl{\"u} and \.{I}\c{s}can(2014)]{bizimpaper}
G{\"u}nl{\"u}, O.; \.{I}\c{s}can, O.
\newblock {DCT} based ring oscillator physical unclonable functions.
\newblock  {IEEE} Int. Conf. Acoustics, Speech and Sign. Proc.; Florence, Italy, May 2014; pp. 8198--8201.

\bibitem[G{\"u}nl{\"u} \em{et~al.}(2015)G{\"u}nl{\"u}, \.{I}\c{s}can, and
Kramer]{bizimtemperature}
G{\"u}nl{\"u}, O.; \.{I}\c{s}can, O.; Kramer, G.
\newblock Reliable secret key generation from physical unclonable functions
under varying environmental conditions.
\newblock  IEEE Int. Workshop Inf. Forensics and Security; Rome, Italy, Nov. 2015; pp. 1--6.



\bibitem[Ignatenko and Willems(2010)]{IgnaFuzzy}
Ignatenko, T.; Willems, F.M.
\newblock Information leakage in fuzzy commitment schemes.
\newblock {\em IEEE Trans. Inf. Forensics and Sec.} June {\bf 2010}, {\em
	5},~2337--348.

\bibitem[Ignatenko and Willems(2014)]{ignatenko2014privacy}
Ignatenko, T.; Willems, F.M.
\newblock Privacy-leakage codes for biometric authentication systems.
\newblock  IEEE Int. Conf. Acoustics, Speech and Sign. Proc.; Florence, Italy, May 2014; pp.
1601--1605.

\bibitem[Maes \em{et~al.}(2012)Maes, Herrewege, and Verbauwhede]{Pufky}
Maes, R.; Herrewege, A.V.; Verbauwhede, I.
\newblock {PUFKY}: A fully functional {PUF}-based cryptographic key generator.
In {\em Cryptographic Hardware and Embedded Sys.}; Berlin Heidelberg,
Germany: Springer-Verlag, Sep. 2012; pp. 302--319.

\bibitem[Maiti and Schaumont(2011)]{maiti2011improved}
Maiti, A.; Schaumont, P.
\newblock Improved ring oscillator {PUF}: an {FPGA}-friendly secure primitive.
\newblock {\em J. Cryptology} Apr. {\bf 2011}, {\em 24},~375--397.

\bibitem[Ahlswede and Csisz{\'a}r(1993)]{AhlswedeCsiz}
Ahlswede, R.; Csisz{\'a}r, I.
\newblock Common randomness in information theory and cryptography - {P}art
{I}: Secret sharing.
\newblock {\em IEEE Trans. Inf. Theory} July {\bf 1993}, {\em 39},~1121--1132.

\bibitem[Maurer(1993)]{Maurer}
Maurer, U.
\newblock Secret key agreement by public discussion from common information.
\newblock {\em IEEE Trans. Inf. Theory} May {\bf 1993}, {\em 39},~2733--742.



\bibitem[Eiroa and Baturone(2011)]{ROAnalysis}
Eiroa, S.; Baturone, I.
\newblock An analysis of ring oscillator {PUF} behavior on {FPGAs}.
\newblock  IEEE Int. Conf. Field-Program. Techn.; New Delhi, India, Dec. 2011; pp. 1--4.

\bibitem[Maiti \em{et~al.}(2010)Maiti et~al.]{ROLarge}
Maiti, A.; others.
\newblock A large scale characterization of {RO-PUF}.
\newblock  IEEE Int. Symp. Hardware-Orient. Sec. and Trust; Anaheim, {CA}, June 2010; pp.
94--99.

\bibitem[Sugiura(1978)]{CorrAIC}
Sugiura, N.
\newblock Further analysis of the data by {A}kaike's information criterion and
the finite corrections.
\newblock {\em Commun. Statistics, Theory and Methods} Jan. {\bf 1978}, {\em
	7},~13--26.

\bibitem[Schwarz(1978)]{BIC2}
Schwarz, G.
\newblock Estimating the dimension of a model.
\newblock {\em The {A}nnals of Stat.} {\bf 1978}, {\em 6},~461--464.

\bibitem[Wang(2012)]{mySPbook}
Wang, R.
\newblock {\em Introduction to orthogonal transforms: with applications in data
	processing and analysis}; Cambridge University Press,  2012.

\bibitem[Bishop(2006)]{MLE}
Bishop, C.M.
\newblock {\em Pattern recognition and machine learning}; Vol.~1, New York:
Springer-Verlag,  2006.

\bibitem[Ohm(2015)]{decorrelation}
Ohm, J.R.
\newblock {\em Multimedia signal coding and transmission}; Berlin Heidelberg,
Germany: Springer-Verlag,  2015.

\bibitem[Puchinger \em{et~al.}(2015)Puchinger et~al.]{Ulm}
Puchinger, S.; others.
\newblock On error correction for physical unclonable functions.
\newblock  {VDE} Int. ITG Conf. Systems, Comm. and Coding; Hamburg, Germany, Feb. 2015; pp. 1--6.

\bibitem[Yin and Qu(2013)]{Yin2013improving}
Yin, C.E.; Qu, G.
\newblock Improving {PUF} security with regression-based distiller.
\newblock  {ACM/IEEE} Design Automation Conf.; Austin, {TX}, May 2013; pp. 1--6.

\bibitem[Komatsu and Sezaki(2001)]{DWHTnomultiplication}
Komatsu, K.; Sezaki, K.
\newblock Lossless {2D} discrete {W}alsh-{H}adamard transform.
\newblock  {IEEE} Int. Conf. Acoustics, Speech and Sign. Proc.; Salt Lake City, {UT}, May 2001;
Vol.~3, pp. 1917--1920.


\bibitem{axi4}
\newblock {AMBA} AXI and ACE Protocol Specification AXI3, AXI4, AXI5, ACE and ACE5.
\newblock Dec. 2017, [Online]. Available:\\ developer.arm.com/docs/ihi0022/latest/amba-axi-and-ace-protocol-specification-axi3-axi4-axi5-ace-and-ace5.

\bibitem{axiS}
\newblock {AMBA} {AXI4-Stream} Protocol Specification
v1.0.
\newblock Mar. 2010, [Online]. Available: https://developer.arm.com/docs/ihi0051/latest/amba-axi4-stream-protocol-specification-v10.

\bibitem[Sahoo \em{et~al.}(2013)]{ref1}
Sahoo, D.P.; Mukhopadhyay, D.; Chakraborty, R.S.
\newblock Design of low area-overhead ring oscillator {PUF} with large challenge space.
\newblock  Int. Conf. Reconfigurable Computing FPGAs, Cancun, Mexico,  Dec. 2013; pp. 9--11.
pp. 1--6.

\bibitem[Parrilla \em{et~al.}(2016)]{ref2}
Parrilla, L.; Castillo, E.; Morales, D.P.; Garc{\'\i}a, A.
\newblock Hardware activation by means of {PUFs} and elliptic curve cryptography in field-programmable devices.
\newblock {\em Electronics} Jan. {\bf 2016}, {\em 5}.

\bibitem[Rukhin \em{et~al.}(2001)Rukhin et~al.]{NIST}
Rukhin, A.; others.
\newblock A statistical test suite for random and pseudorandom number
generators for cryptographic applications.
\newblock Technical report, {National Inst. Stand. and Techno.},  2001.
\newblock {Rev.} in 2010.

\bibitem[Lin and Costello(2004)]{ECC}
Lin, S.; Costello, D.J.
\newblock {\em Error control coding}; Englewood Cliffs, NJ: Prentice-Hall,
2004.

\bibitem[Chen and Pratt(1984)]{Zigzag}
Chen, W.H.; Pratt, W.
\newblock Scene adaptive coder.
\newblock {\em IEEE Trans. Commun.} Mar. {\bf 1984}, {\em 32},~225--232.

\bibitem[Hong(2011)]{DFTCF}
Hong, Y.
\newblock On computing the distribution function for the sum of independent and
nonidentical random indicators.
\newblock Technical report, Dep. Stat., Virginia Tech., Blacksburg, VA, Apr. 2011.

\bibitem[Wyner and Ziv(1973)]{WZ}
Wyner, A.D.; Ziv, J.
\newblock A theorem on the entropy of certain binary sequences and applications: {P}art {I}.
\newblock {\em {IEEE} Trans. Inf. Theory} Nov. {\bf 1973}, {\em 19},~769--772.

\bibitem[Polyanskiy \em{et~al.}(2010)Polyanskiy, Poor, and
Verd{\'u}]{Polyanskiy}
Polyanskiy, Y.; Poor, H.V.; Verd{\'u}, S.
\newblock Channel coding rate in the finite blocklength regime.
\newblock {\em {IEEE} Trans. Inf. Theory} May {\bf 2010}, {\em 56},~2307--2359.


\end{thebibliography}

\end{document}